\documentclass[11pt,a4paper]{article}
\usepackage{jheppub,adjustbox,multirow,amsmath}
\usepackage{graphicx,bbold}

\def\hbar{\hspace{0pt}\raisebox{1pt}{$-$} \hspace{-7pt} h}

\def\5{\overline 5}

\newcommand{\beq}{\begin{equation}}
\newcommand{\eeq}{\end{equation}}
\newcommand{\bea}{\begin{eqnarray}}
\newcommand{\eea}{\end{eqnarray}}

\title{An electroweak basis for neutrinoless double $\beta$ decay}
\date{\today
}
\author{Michael L. Graesser}
\affiliation{Theoretical Division T-2, Los Alamos National Laboratory \\ Los Alamos, NM 87545, USA}
\abstract{A discovery of neutrinoless double-$\beta$ decay would be profound, providing the first direct experimental evidence of $\Delta L=2$ lepton number violating processes. While a natural explanation is provided by an effective Majorana neutrino mass, other new physics interpretations should be carefully evaluated. 
At low--energies such new physics could manifest itself in 
the form of color and $SU(2)_L \times U(1)_{Y}$ invariant higher dimension operators. Here we determine a complete set of electroweak invariant dimension--9 operators, and our analysis supersedes those that only impose $U(1)_{em}$ invariance. 
Imposing electroweak invariance implies: 1) a significantly reduced set of leading order operators compared to only imposing $U(1)_{em}$ invariance; and 2) other collider signatures. Prior to imposing electroweak invariance we find a minimal basis of 24 dimension-9 operators, which is reduced to 11 electroweak invariant operators at leading order in the expansion in the Higgs vacuum expectation value. We set up a systematic analysis of the hadronic realization of the 4-quark operators using 
chiral perturbation theory, and apply it to determine which of these operators have long-distance pion enhancements at leading order in the chiral expansion. 
We also find at dimension--11 and dimension--13 the electroweak invariant operators that after electroweak symmetry breaking produce the remaining $\Delta L=2$ operators that would appear at dimension--9 if only $U(1)_{em}$ is imposed. 
}

\begin{document}
\maketitle
%

\section{Introduction}

The existence of neutrino masses and of dark matter both point to new physics beyond the Standard Model. If neutrinos have a Majorana mass, then they would be the only known fundamental fermonic particle that is also its own antiparticle. This would imply that overall lepton number is violated in the vacuum, which could have implications for the origins of the baryon asymmetry, as well as impact astrophysics, such as supernova neutrino oscillations and the r-process. To date the best experimental approach for distinguishing whether neutrinos are Majorana or Dirac is to search for the so-called neutrinoless double beta decay processes, in which 
  \beq 
(A, Z) \rightarrow (A, Z+2) + e^-e^-  ~.
\eeq 
The current best constraints on these processes are from the GERDA and KamLAND-Zen experiments, which have sets limits of $T^{0 \nu}_{1/2} > 2.1 \times 10^{25}$ years for 
$Ge^{76}$ \cite{Agostini:2013mzu} 
and  $T^{0\nu}_{1/2} >  1.07\times10^{26}$~years  for $^{136}$Xe~\cite{KamLAND-Zen:2016pfg}, respectively. These bounds translate to a limit on the effective neutrino mass matrix element that is just above the top of the ``inverted" neutrino mass spectrum. The next-generation of multi-tonne experiments are expected to reach sensitivities that extend to the bottom of the inverted neutrino mass hierarchy spectrum, while remaining insensitive to Majorana neutrino masses having a normal hierarchy. 

Given the significance of a discovery of a $\Delta L=2$ process, exploring alternative interpretations of a {\em positive} neutrinoless double beta signal requires some urgency to avoid, in the face of a positive signal, making the wrong inference about the size of the effective Majorana neutrino mass.  Inferring the effective Majorana neutrino mass from the observed lifetime is a step that would require independent evidence. For such a signal could be due to the exchange of some new exotic particles at short distances (see for instance, the review \cite{Rodejohann:2011mu}), rather than the effective neutrino mass of Majorana neutrinos. If so, other experiments will be required to sort out a large number of degeneracies in the space of theoretical possibilities. In the circumstance that neutrino masses have a normal hierarchy (established by, for example, short-baseline experiments) and a positive neutrinoless double beta rate is observed, exotic interpretations would be inevitable. 

For neutrinoless double beta decay experiments, one approach for resolving degeneracies is to obtain more information about each event. 
Future experiments, such as NEXT \cite{Gomez-Cadenas:2014dxa} and SuperNemo \cite{Pahlka:2008dw}, plan to measure the individual energies of the two electrons and their relative separation angle. 
With this additional kinematic information, 
forward-backward like correlations of the separation angle or energy have the potential to distinguish a signal arising from a Standard Model (SM) long-distance neutrino exchange from that arising from a short-distance 
process \cite{Doi:1982dn, Doi:1985dx, Tomoda:1986yz, Ali:2007ec, Arnold:2010tu,Horoi:2015gdv}. 
Another experiment that has a potential to resolve degeneracies in exotic explanations would be the Large Hadron Collider (LHC) or a future hadron collider, which might be able to directly probe the scale of the new physics. Ref. \cite{Helo:2013ika} discusses using a charge asymmetry
or invariant mass peaks to resolve degeneracies between specific models.

A novel short distance contribution to a neutrinoless double beta decay signal 
appears at low-energy in the form of  $\Delta L=2$ violating higher dimension operators of the type 
\beq
{\cal L} = \frac{1}{\Lambda^5} \sum_i c_i {\cal O}_i ,
\label{general-L} 
\eeq
\beq
{\cal O}_i  \sim  \left( \overline{u} \Gamma d) (\overline{u} \Gamma^\prime d) (\overline{e} \Gamma ^{\prime \prime} e^c \right),
\label{op:general-dim9}
\eeq 
for some Dirac matrices $\Gamma, \Gamma^\prime, \Gamma^{\prime \prime}$. \footnote{Operators in which a quark and a lepton are in the same bilinear can be eliminated by a Fierz transformation. The interested reader is referred to Section \ref{Appendix:complete-set} for more details.} Because of the high dimensionality of this operator, the neutrinoless beta decay rate $1/T_{1/2} \propto \Lambda^{-10}$ is easily suppressed. Still, neutrinoless beta decay experiments are currently probing the multi-TeV region, which is of considerable interest given that the LHC is probing the same scale. 

In a neutrinoless double beta decay process two neutrons inside a nucleus ``collide" to (very rarely) produce two protons and two electrons with no neutrinos. It is well-known that the same process can be searched for at a hadron collider experiment,  where such short-distance operators contribute to \cite{Keung:1983uu}
\begin{itemize} 
\item same-signed (SS) dilepton process,  $ p p \rightarrow \ell ^\pm \ell ^\pm +2j$.
\end{itemize} 
Using this signature a number of constraints have been proposed or obtained on specific models using then forthcoming \cite{Tello:2010am} or actual 8 TeV LHC data \cite{Nemevsek:2011hz,
Helo:2013dla, Helo:2013ika, Deppisch:2015qwa}. Projections for future LHC sensitivities at 13 TeV center-of-mass energy and with O(100 fb$^{-1}$) of integrated luminosity indicate that for specific models the LHC will be competitive with existing GERDA bounds and future 1 tonne experiments \cite{Helo:2013ika, Peng:2015haa, Deppisch:2015qwa}.
\footnote{To obtain a reliable comparison between future LHC and next generation neutrinoless double beta experiments, Ref. \cite{Peng:2015haa} improves on the results of Ref. \cite{Helo:2013ika} in several ways. They include the QCD running of operators between the TeV and GeV scales, the important long-distance pion contribution to the nuclear matrix element, 
and for the collider analysis, include backgrounds and a detector simulation.}
 Current and future LHC searches are competitive with neutrinoless double beta experiments as they benefit from the enhancement in the production cross-section due to the on-shell production of the mediators of the higher dimension operators in Eqs.~(\ref{op:general-dim9}). \footnote{The selection efficiency for the signal has a dramatic dependence on the mass of the intermediate particles, as previously shown in the context of using monojet searches at the LHC and Tevatron to bound non-standard neutrino interactions \cite{Friedland:2011za}. At low and high mass the efficiency drops: at low mass, since for fixed analysis cuts the $p_T$ spectrum is falling, and at high mass as the contact limit is approached the available phase space in the resonance channels decreases.}
 In the contact limit, collider cross-sections and neutrinoless double beta decay rates have the same scaling with $\Lambda$, so the gain in sensitivity of one experiment compared to the other is linear. 

One of the motivations for this present work is based on the simple observation that in general the operators appearing above in Eq.~(\ref{op:general-dim9}) are not gauge-invariant under the full electroweak symmetry of the Standard Model. This work takes the next natural step of generalizing the operators 
in Eq.~(\ref{op:general-dim9}) to their full SM invariance. While previous work \cite{Babu:2001ex} classifies operators by their SM gauge invariance, but omits the last step of classifying operators by their Lorentz structure, other work \cite{Pas:2000vn, Prezeau:2003xn} classifies operators by their Lorentz structure and $SU(3)_c \times U(1)_{em}$ invariance, but not by their electroweak gauge invariance. Here these previous results are extended in two directions: i) by presenting a minimal basis of operators classified by both their electric and color invariance and by their Lorentz structure; and ii)
by presenting a minimal basis of operators -- a subset of the previous set -- classified by both their SM gauge invariance and Lorentz structure.  On point i), the present works corrects the literature \cite{Pas:2000vn, Prezeau:2003xn} on a minimal set of electromagnetic and color invariant operators. Compared to Ref. \cite{Prezeau:2003xn} we find additional operators that differ
on the way color is contracted among the 4-quarks, namely that so-called ``color-octet'' operators should be included. The ``super-formula'' of Ref. \cite{Pas:2000vn} similarly does not have the color-octet operators, and as previously noted in \cite{Prezeau:2003xn}, has an extra tensor operator involving the two electrons that can be eliminated. The present work presents a minimal 
basis of color and electromagnetic invariant operators that can be used as a starting point for relating neutrinoless double beta decay observables to models and observables defined at 
a higher mass scale.

But why complete such operators to their $SU(2)_L \times U(1)_Y$ invariant form? 
This effective field theory approach has several obvious benefits 
when the mass scale of the $\Delta L=2$ physics scenarios is much larger than the electroweak scale, which shall be assumed throughout. At a general level:
\begin{itemize} 
\item To determine the effect of a specific model on the neutrinoless double beta decay rate one has to simply match the model onto the Wilson coefficients of the effective theory that is the SM plus a minimal set of $\Delta L=2$ electroweak invariant operators. The mixing and evolution of these operators due to QCD and electroweak interactions is then simply described by standard renormalization group techniques. The universality of the renormalization group evolution is separated from the details of the model.
These operators are then matched to operators in the chiral theory of nucleons and meson at the QCD scale, using inputs from lattice QCD. All of this is standard practice in, for instance, determining the effects of new physics scenarios on $K$ and $B$ meson physics. The only model-dependent input is in the matching of the Wilson coefficients at the high scale; the rest is universal.
\item When the particles that resolve the low-energy $\Delta L=2$ operators are too heavy to be a produced at a collider, then the collider experiment is only probing the contact operator. In this limit the electroweak invariant effective theory provides a universal intermediate effective theory for direct apples-to-apples comparisons between low-energy and collider experiments.
\end{itemize} 
An alternative approach is to consider all possible models that at low-energy realize color and electromagnetic invariant $\Delta L=2$ operators, such as done in \cite{Bonnet:2012kh}. As discussed previously, collider signatures will depend on the model when the intermediate particles can be produced on-shell. However, in the contact limit, the ``all models'' approach has to reduce to the electroweak invariant effective field theory. At energies below the mass of the new particles, all the dependence of any model is subsumed into the Wilson coefficients of the electroweak invariant effective field theory.

At a more specific level, there are additional benefits and findings:
\begin{itemize}
\item As mentioned above, organizing operators by their $SU(2)_L \times U(1)_Y$ invariance provides a useful basis for studying mixing among these operators due to QCD and especially electroweak interactions. Since the electroweak interactions violate parity, mixing of operators due to electroweak renormalization may lead to an important effect at low-energies. Namely, it could cause an operator that otherwise appears at a higher order in the chiral power counting -- because of parity -- to appear at a lower chiral order. Whether such an effect occurs or not is an open question.
\item While in the $SU(3)_c \times U(1)_{em}$ effective theory we find a minimal basis of 24 baryon conserving, $\Delta L=2$ operators that contribute at leading order in $1/\Lambda$ (which turns out to be dimension-9), in the $SU(3)_c \times  SU(2)_L \times U(1)_Y$ invariant effective theory at leading order in $v/\Lambda$ one instead finds only a subset of operators. The reason is simple: only 11 of the operators in Eqn. (\ref{op:general-dim9}) conserve $U(1)_Y$ hyper-charge. To conserve hyper-charge in these other operators one has to go to higher dimensions by inserting powers of the Higgs field $H$.  In particular, at dimension--11 one can insert two Higgs fields and one finds another 12 out of the 24 operators;  one needs to go to dimension 13 -- requiring 4 Higgs insertions -- to obtain all the low-energy dimension-9 operators. 
\item Below the weak scale one therefore expects at most 11 of the 24 operators to be phenomenologically relevant. For 7 of these operators, the 4-quark part of the operator is scalar, and for the other 4 operators it is vector. Specific models may generate the other operators in the UV, but at low energies those are suppressed by at least $v^2/\Lambda^2$. 
\item  
The electroweak completion of a given low-energy operator may imply additional channels to search for these operators at hadron colliders. 
As previously noted, such operators produce same-signed dilepton signals. But the requirement of electroweak invariance may imply additional final states in which to search for such $\Delta L=2$ operators; whether this occurs is specific to that operator. 
If one or more leptons in the SS dilepton final state are left-handed, then an $SU(2)_L$ rotation can turn it into a neutrino. So in addition to SS dileptons, 
we can also expect to find
\begin{itemize} 
\item lepton + MET final states, $ p p \rightarrow \ell ^\pm+ 2j$+MET,
\item MET + multi jet final states, $ p p \rightarrow 2j $+MET,
\end{itemize} 
all occurring at comparable rates. 
While in practice SM backgrounds are significantly smaller for the SS final state, 
these other channels could be used, at least in principle, to test competing hypotheses for a $\Delta L=2$ process. For example, \cite{Helo:2015ffa} uses 8 TeV LHC dijet data to constrain 
specific models, such as the left-right symmetric model, or models that involve leptoquarks or charged scalars, and 
 \cite{Helo:2015ffa, Gonzales:2016krw} present projections of the sensitivity of future dijet and leptoquark searches to such models.
\end{itemize}

The second motivation for the present work is the following.
To obtain predictions for the neutrinoless double $\beta$ decay rate from the effective Lagrangian in Eq. (\ref{general-L}), one needs at an intermediate step the matrix elements of the 4-quark operators appearing in Eq. (\ref{op:general-dim9}) between external pions and nucleons. 
The most important chiral interactions and Feynman diagrams are shown in Fig. \ref{fig:chiPT}.
With the matrix elements as input, one then uses chiral perturbation theory to obtain amplitudes (or potentials) at the nucleon level, which are then used as inputs into nuclear structure computations. 
The 4-quark operators lead to a number of operators in the chiral effective theory, each with a low-energy constant, which will all eventually be computed using
using lattice QCD. 

In what follows the values of the low-energy constants will not be needed. However, in the chiral effective field theory, the most important coupling of the leptons induced by the interactions in Eq. (\ref{general-L}) is with two pions -- if it exists -- rather than a direct contact interaction with four nucleons. The reason is that in the chiral effective theory, the amplitude for a neutrinoless double decay process arising from the two-pion coupling -- diagram (a) in Fig. \ref{fig:chiPT} -- is chirally enhanced compared to that caused by a direct four nucleon interaction. While 
the importance of a long-distance pion contribution was noticed in supersymmetric models of $R-$parity violation some time ago \cite{Faessler:1996ph}, much of the literature continues to ignore 
the two-pion coupling and instead assumes the nuclear matrix element to be given by the direct 4-nucleon coupling (i.e., diagram (c) in Fig. \ref{fig:chiPT}). However, by now the two-pion coupling cannot be glossed over.  Preliminary results for the matrix elements of the operators between two pions are now available \cite{Nicholson:2016byl}. And from using chiral $SU(3)$, the same $\pi \pi$ matrix elements can be estimated at the $O(30\%)$ level from 
$K \pi \pi$ \cite{Savage:1998yh} \cite{Cirigliano:2017ymo} and
$K^0$-$\overline{K}^0$ \cite{Cirigliano:2017ymo} matrix elements, which have been computed using lattice QCD.

Which quark operators lead to two-pion interactions is clearly important for the phenomenology of neutrinoless double beta decay, and this question was broadly investigated in \cite{Prezeau:2003xn}. One surprising finding from Ref. \cite{Prezeau:2003xn} is that that not all operators appearing in Eqs. (\ref{general-L}) and (\ref{op:general-dim9}) lead to $\pi \pi$ interactions at leading order in the chiral expansion. If true, then for those operators, the neutrinoless double $\beta$ decay rate may not be chirally enhanced, and the collider constraints more competitive. 

Here we revisit this analysis, and set up a systematic matching of the operators appearing in Eq. (\ref{op:general-dim9}) to operators in the chiral theory. We show in general how to determine the most important chiral operators, which are the interactions of two electrons to two pions, two nucleons and a pion, and four nucleons. We work out all the leading chiral order interactions in detail of two electrons with two pions, reproducing to next-to-next lowest order (NNLO) the power counting
results of Ref. \cite{Prezeau:2003xn}. 
Compared to that reference, here we find more chiral operators, each of which at low-energies appears with its own low-energy constant.

Here is the outline. In the sections that follow we first present a minimal basis of operators that at low-energy contribute to neutrinoless double $\beta $ decay. This minimal basis is 
derived in some detail in Section \ref{Appendix:complete-set}. At the level of only color and electromagnetic invariant $\Delta L=2$ dimension-9 operators, we find a larger minimal basis of operators -- 24 -- compared to Ref. \cite{Prezeau:2003xn}, which finds 14 operators. \footnote{They find 5 scalar 4-quark operators and 4 vector 4-quark operators, leading to 14 independent dimension-9 operators. Compared to that reference, here we include color-octet operators which cannot be eliminated by color or Dirac Fierzing, and this adds 3 more scalar operators and 4 more vector operators.}
Since such operators are only $SU(3)_c \times U(1)_{\rm em}$ invariant, in the subsequent sections we make such operators $SU(2)_L \times U(1)_Y$ gauge invariant; at low-energies these operators map into a subset of the complete set of neutrinoless double $\beta$ operators. We discuss dimension-9, -11, and -13 electroweak invariant operators in sections \ref{sec:Dimension-9}, \ref{sec:Dimension-11} and \ref{sec:Dimension-13}. In Section \ref{sec:chiralPT} we set up the matching of the 4-quark operators onto operators defined in the chiral theory, and work out in some detail the chiral two pion operators. We then conclude in Section \ref{conclusions}.

\section{Below the electroweak and $\Delta L=2$ mass scales} 

Below both the electroweak scale and the mass scale ($\Lambda$) of the new $\Delta L=2$ physics, the physics of the lepton number violating processes is described by a series of $\Delta L=2$ violating higher dimension operators.  The leading operators that contribute at short-distances to a neutrinoless double $\beta$ decay signal involve 4 quarks and 2 charged leptons and are dimension-9. \footnote{We do not consider here $\Delta L=2$ operators involving field strength tensors, covariant derivatives, those that are anti-symmetric in the lepton flavor indices, or those that involve an electron and a neutrino in the final state, instead of two electrons. See Ref. \cite{Babu:2001ex} for a more general set of possibilities. \label{footnote-qualifier}}
At these low energies such operators must be explicitly $SU(3)_c \times U(1)_{\rm em}$ invariant. To leading order in $1/\Lambda$, we find that a minimal basis of such $\Delta L=2$, $B$ conserving operators is given by 
\begin{eqnarray}
{\cal L}_{eff} &=&
\frac{1}{\Lambda^{5}} \left[\sum_{i=scalar} (c^s_i  \overline{e} e^c + c^{\prime s}_i  \overline{e} \gamma^5 e^c ){\cal O}_{s,i}
+ \overline{e} \gamma_\mu \gamma^ 5 e^c  \sum_{i=vector} c^v_i 
{\cal O}^{\mu}_{v,i} \right]
\label{4q2l-op}
\end{eqnarray}
where the sum is over the set of scalar 4-quark operators $\{{\cal O}_{s,i}\}$ and set of vector 4-quark operators $\{{\cal O}^{\mu}_{v,i}\}$.  

The following basis of quark operators is convenient in order to classify the hadronic realization of these 4-quark operators using their transformation properties under chiral $SU(2)_L \times SU(2)_R$.  
Dropping the subscripts $s$ and $v$, we find the following 8 scalar operators
\begin{subequations}
\bea
{\cal O}_{1LR} & =& (\overline{q}_L \gamma^\mu \tau^+ q_L) (\overline{q}_R \gamma_\mu \tau^+ q_R), 
\label{scalarO1LR} \\
{\cal O}^\lambda_{1LR}  &=& (\overline{q}_L \gamma^\mu \tau^+ \lambda^Aq_L) (\overline{q}_R \gamma_\mu \tau^+ \lambda^A q_R), 
\label{scalarOl1LR} \\ \\
{\cal O}_{2RL} &=& (\overline{q}_R  \tau^+ q_L) (\overline{q}_R  \tau^+ q_L), 
\label{scalarO2RL} \\
{\cal O}^\lambda_{2RL} &=& (\overline{q}_R  \tau^+\lambda^A q_L) (\overline{q}_R  \tau^+ \lambda^A q_L), 
\label{scalarOl2RL} \\ \\
{\cal O}_{2LR} &=&(\overline{q}_L  \tau^+ q_R) (\overline{q}_L  \tau^+ q_R), 
\label{scalarO2LR} \\ 
{\cal O}^\lambda_{2LR} &=&(\overline{q}_L  \tau^+ \lambda^A q_R) (\overline{q}_L  \tau^+ \lambda^A q_R), 
\label{scalarOl2LR} \\ \\
{\cal O}_{3L} &=& (\overline{q}_L \gamma^\mu \tau^+ q_L) (\overline{q}_L\gamma_\mu \tau^+ q_L) , 
\label{scalarO3L}\\ \\
{\cal O}_{3R} &=&(\overline{q}_R \gamma^\mu \tau^+ q_R) (\overline{q}_R\gamma_\mu \tau^+ q_R), 
\label{scalarO3R}
\eea
\end{subequations}
and 8 vector operators
\begin{subequations}
\bea
{\cal O}^\mu_{LLLR} &=& (\overline{q}_L \gamma^\mu \tau^+ q_L)(\overline{q}_L \tau^+ q_R), 
\label{vectorOLLLR} \\
{\cal O}^{\lambda,\mu}_{LLLR} &=& (\overline{q}_L \gamma^\mu \tau^+ \lambda^A q_L)(\overline{q}_L \tau^+ \lambda^A q_R), 
\label{vectorOlLLLR} \\ \nonumber \\
{\cal O}^\mu_{RRLR} &=& (\overline{q}_R \gamma^\mu \tau^+ q_R)(\overline{q}_L \tau^+ q_R), 
\label{vectorORRLR}  \\
{\cal O}^{\lambda,\mu}_{RRLR} &=& (\overline{q}_R \gamma^\mu \tau^+ \lambda^A q_R)(\overline{q}_L \tau^+ \lambda^A q_R), 
\label{vectorOlRRLR} \\ \nonumber \\
{\cal O}^\mu_{LLRL} &=& (\overline{q}_L \gamma^\mu \tau^+ q_L)(\overline{q}_R \tau^+ q_L), 
\label{vectorOLLRL} \\
{\cal O}^{\lambda,\mu}_{LLRL} &=& (\overline{q}_L \gamma^\mu \tau^+ \lambda^A q_L)(\overline{q}_R \tau^+ \lambda^A q_L), 
\label{vectorOlLLRL} \\ \nonumber \\
{\cal O}^\mu_{RRRL} &=& (\overline{q}_R \gamma^\mu \tau^+ q_R)(\overline{q}_R \tau^+ q_L), 
\label{vectorORRRL} \\ 
{\cal O}^{\lambda,\mu}_{RRRL} &=& (\overline{q}_R \gamma^\mu \tau^+ \lambda^A q_R)(\overline{q}_R \tau^+ \lambda^A q_L), 
\label{vectorOlRRRL} 
\eea
\end{subequations}
where $q_{L/R}=(u~d)_{L/R}$, and $\tau^+ = \left( \begin{array}{cc} 
	0 & 1 \\ 
	0 & 0 \end{array} \right)$ . $\lambda^A$, $A$=1.,..$8$, refer to the $SU(3)$ color generators in the fundamental representation,  
	and implicit summation over $A$ is assumed. The relation of these operators to those defined in Ref. \cite{Prezeau:2003xn} is given in Section \ref{app:relationtootherdefs}.
In total 24 different 6-fermion operators, involving 16 different 4-quark operators, can appear in the Lagrangian, each with its own Wilson coefficient $\{c^s_i, c^{\prime s}_i, c^v_i\}$. All quark bilinears can be arranged to be either color-singlets or color-octets. Operators not appearing in this set either vanish, or can be reduced to a linear combination of the operators in this set through color Fierz and/or generalized Fierz transformations, as we show in the Appendix (Section \ref{Appendix:complete-set}). 
We note that these 8 scalar operators are 
equivalent to the basis presented in \cite{Buras:2000if}. \footnote{A previous version of this manuscript presented a larger minimal basis of 10 scalar operators equivalent (after some Fierzing) to the 10 4-quark operators used in analyses of beyond-the-Standard Model contributions to $\Delta S=2$ processes \cite{Bagger:1997gg}. The author thanks V. Cirigliano,  W. Dekens, E. Mereghetti and B. Tiburzi for discussions on reducing the operator basis through eliminating 4-quark operators of the form $\sigma^{\mu \nu} \otimes \sigma_{\mu \nu}$. The author finds that all of the vector operators of the form $\gamma _\nu \otimes \sigma^{\nu \mu}$ appearing in a previous version of the manuscript can be removed by Fierz identities.} 
Operators in groups separated by line breaks have the same chiral structure and will mix with each other under QCD renormalization.

\section{Below the $\Delta L=2$ mass scale: weak scale operators} 

The physics of the $\Delta L=2$ processes is assumed to be higher than the weak scale, so that at scales below $\Lambda$ such physics can be characterized by a series of higher dimension operators expanding in $1/\Lambda$ and $v/\Lambda$. Above the weak scale we are assuming the electroweak symmetry is linearly realized with a single Higgs boson doublet $H$. In this section we determine {\em all} the lowest dimension $\Delta L=2$ (and baryon conserving) operators that are invariant under the SM gauge symmetry, that after electroweak symmetry breaking give operators found in Eqn. (\ref{4q2l-op})
\footnote{See footnote \ref{footnote-qualifier}.}. 
As we shall see, such operators form a subset of the operators appearing in Eqn. (\ref{4q2l-op}).

Our notation is the following: $Q=(u ~d)_L$, $\ell=(\nu~ l)_L$, and $H$ is the Higgs doublet of the SM with hypercharge assigment $+1/2$ with vacuum expectation value (vev) $H \rightarrow (0 ~ v/\sqrt{2})$, $v\simeq247 $ GeV;
Roman letters $a, b, c,...=1,2$ refer to $SU(2)_L$ indices.  We form $SU(2)_L$ invariants using $\delta^a_b$ and $\epsilon=i \sigma^{(2)}$ (with $\epsilon_{12}=+1$), and use 
the $SU(2)_L$ Fierz identity to eliminate $\sigma^a \sigma^a$ in favor of $\delta$'s.  
We only consider operators involving first generation fields, since our focus is on those operators which contribute directly to neutrinoless double $\beta$ decay. 

\subsection{Dimension--9} 
\label{sec:Dimension-9}

These operators necessarily involve 4 quark fields and 2 lepton fields, and therefore do not involve any Higgs fields. We organize the operators by whether the lepton bilinear is $\sim \overline{\ell} \ell^C$, $\overline{e}_R e^C_R$ or $\overline{\ell} e^C_R$. 

\subsubsection{$\overline{\ell} \ell^C$}
The operators in this category involve two lepton doublets and therefore the number of quark doublets must be even.  Hypercharge isn't conserved with zero or four quark doublets. 
That leaves three operators containing two quark doublets, corresponding to three possibilities for the remaining two (right-handed) quarks: $\overline{u}_R \overline{u}_R$, $\overline{u}_R d_R$, and
$d_R d_R$. To obtain $\overline{u} d$, each quark doublet has to $SU(2)_L$ contract with a lepton doublet, since they can't contract with each other: $\overline{Q}Q=\overline{u}_Lu_L+\overline{d}_Ld_L$ and $QQ=0$ if the two $Q$'s are from the same generation. If two $Q$'s are from the different generation, then $Q Q^\prime$ can be non-zero, but then the two lepton doublets would have to $SU(2)_L$ contract with each other and that would require lepton doublets from two different generations, a possibility we do not explore here. The electroweak contractions are unique. 
\begin{align*}
{\rm LM1} &=  i \sigma^{(2)}_{a b} (\overline{Q}_{a}  \gamma^\mu Q_{c})
     ( \overline{u}_R  \gamma_\mu d_R) 
     ( \overline{\ell}_b  \ell^C _c ) \\ 
     & =  (\overline{u}_R \gamma^\mu d_R) \Big[ (\overline{u}_L \gamma_\mu d_L) (\overline{e}_L e^C_L) 
       +(\overline{u}_L \gamma_\mu u_L - \overline{d}_L \gamma_\mu d_L )(\overline{e}_L \nu_L^C)  \nonumber \\ 
   &    ~~~~~~~~~~~~~~~~ - (\overline{d}_L \gamma_\mu u_L )(\overline{\nu}_L \nu_L^C) \Big] \\
   {\rm LM2} &=  i \sigma^{(2)}_{a b} (\overline{Q}_{a}  \gamma^\mu \lambda^A Q_{c})
     ( \overline{u}_R  \gamma_\mu \lambda^A d_R) 
     ( \overline{\ell}_b  \ell^C _c ) \\ 
{\rm LM3} &=    (\overline{u}_R Q_{a} ) (\overline{u}_R Q_{b})
     (   \overline{\ell}_{a}  \ell^C_{b}) \\
     &= \Big[ (\overline{u}_R d_L) (\overline{u}_R d_L)  (\overline{e}_L e^C_L) 
     + 2 (\overline{u}_R d_L) (\overline{u}_R u_L) (\overline{e}_L \nu_L^C ) \nonumber \\
 &  ~~~+ (\overline{u}_R u_L) ( \overline{u}_R u_L) (\overline{\nu}_L \nu_L^C )  \Big]  \\
 {\rm LM4} &=    (\overline{u}_R \lambda^A Q_{a} ) (\overline{u}_R  \lambda^A Q_{b})
     (   \overline{\ell}_{a}  \ell^C_{b}) \\
{\rm LM5} &= i \sigma^{(2)}_{ab} i \sigma^{(2)}_{cd} (\overline{Q}_a d_R) (\overline{Q}_{c} d_R)
       ( \overline{\ell} _{b}  {\ell}^C_d)  \\
       & = \Big[ (\overline{u}_L d_R) (\overline{u}_L d_R) (\overline{e}_L e^C_L) 
       	-2 (\overline{u}_L d_R) (\overline{d}_L d_R) (\overline{e}_L \nu_L^C ) \nonumber \\ 
 &  ~~~+ (\overline{d}_L d_R) ( \overline{d}_L d_R) (\overline{\nu}_L \nu_L^C )  \Big]	\\
 {\rm LM6} &= i \sigma^{(2)}_{ab} i \sigma^{(2)}_{cd} (\overline{Q}_a \lambda^A d_R) (\overline{Q}_{c}  \lambda^A d_R)
       ( \overline{\ell} _{b}  {\ell}^C_d)  
\end{align*}

Here we see that LM1--LM6 operators contribute to all three types of hadron collider signatures.  For 
 the LM1 operator, for example, the individual component operators in the first line contribute to a same-signed di-lepton signal and to MET + a single lepton, and the component operator in the last line to $2j +$ MET. Individually, each LM operator contributes more or less with equal rates to each of these hadron collider signatures.
In LM1, LM3 and LM5, each quark bilinear is a color-singlet, whereas in the operators LM2, LM4 and LM6, each quark bilinear transforms under $SU(3)_c$ as a color-octet.

\subsubsection{$\overline{e}_R e^C_R$}
Next we have our first dimension--9 operator that only contributes to a same--signed dilepton signal, simply because the $SU(2)_L$ invariant operator does not involve any left-handed lepton fields, and hence no neutrinos. It is  
\begin{align*}
{\rm LM7 } &=  (\overline{u}_R \gamma^\mu d_R)
	(\overline{u}_R \gamma_\mu d_R)
        (\overline{e}_R  e^C_R) 
\end{align*}
A Fierz transformation shows this operator is identical to the operator where the color is contracted between quark and anti-quarks of different bilinears. The color-octet operator is therefore not independent from the operator above. 
There are no operators in this sub-category involving quark doublets because 
one can't conserve hypercharge.  

\subsubsection{$\overline{\ell} \gamma^\mu e^C_R$}
The operators in this set contain one right-handed $e_R$ and one left-handed $\ell$ field,
so the operators in this sub-category must have an odd number of quark doublets to obtain an $SU(2)_L$ invariant. 
The 4 quarks must have a total hypercharge of $-3/2$ to cancel that of the leptons, a consideration that excludes the possibility of 
three quark doublets. The two choices below correspond to having a quark doublet or an anti-quark doublet. The electroweak contractions are unique. 
\begin{align*}        
{\rm LM8 } &= ( \overline{u}_R \gamma^\mu d_R)
	 i \sigma^{(2)}_{ab} (\overline{Q}_{a}  d_R)
         ( \overline{\ell}_{b} \gamma_\mu  e^C_R) \\
           &=  (\overline{u}_R \gamma^\mu d_R) 
           \Big[ (\overline{u}_L d_R) (\overline{e}_L \gamma_\mu e^C_R)   - (\overline{d}_L d_R) (\overline{\nu}_L \gamma_\mu e^C_R)  \Big] \\  
{\rm LM9 } &= ( \overline{u}_R \gamma^\mu \lambda^A d_R)
	 i \sigma^{(2)}_{ab} (\overline{Q}_{a}  \lambda^A d_R)
         ( \overline{\ell}_{b} \gamma_\mu  e^C_R) \\
{\rm LM10} &=  (\overline{u}_R \gamma^\mu d_R) 
	(\overline{u}_R Q_a)
      (\overline{\ell}_a \gamma_\mu  e^C_R)     \\
      &= (\overline{u}_R \gamma^\mu d_R) \Big[ (\overline{u}_R d_L) (\overline{e}_L \gamma_\mu e^C_R) 
    + (\overline{u}_R u_L) (\overline{\nu}_L \gamma_\mu e^C_R)  \Big]\\
{\rm LM11} &=  (\overline{u}_R \gamma^\mu \lambda^A d_R) 
	(\overline{u}_R \lambda^A Q_a)
      (\overline{\ell}_a \gamma_\mu  e^C_R) 
\end{align*}
They each contribute to a same-signed dilepton signal as well as to a lepton + MET signal, but not to a $2j+$ MET signal. 

Operators of the form $\gamma_\nu \otimes \sigma^{\nu \mu}$ can be eliminated by generalized Fierz transformations; the reader is referred to Section \ref{Appendix:complete-set} for further details.

\subsubsection{Dimension-9 summary}
In short, at this dimension we have the following set of 15 electroweak invariant operators:
\begin{subequations}
\bea 
{\rm LM1} &=&  i \sigma^{(2)}_{a b} (\overline{Q}_{a}  \gamma^\mu Q_{c})
     ( \overline{u}_R  \gamma_\mu d_R) 
     ( \overline{\ell}_b  \ell^C _c )  \\ 
{\rm LM2} &=&  i \sigma^{(2)}_{a b} (\overline{Q}_{a}  \gamma^\mu \lambda^A Q_{c})
     ( \overline{u}_R  \gamma_\mu \lambda^A d_R) 
     ( \overline{\ell}_b  \ell^C _c ) \\ 
{\rm LM3} &=&    (\overline{u}_R Q_{a} ) (\overline{u}_R Q_{b})
     (   \overline{\ell}_{a}  \ell^C_{b})  \\
 {\rm LM4} &= &   (\overline{u}_R \lambda^A Q_{a} ) (\overline{u}_R  \lambda^A Q_{b})
     (   \overline{\ell}_{a}  \ell^C_{b}) \\     
{\rm LM5} &=&  i \sigma^{(2)}_{ab} i \sigma^{(2)}_{cd} (\overline{Q}_a d_R) (\overline{Q}_{c} d_R)
       ( \overline{\ell} _{b}  {\ell}^C_d)  \\
   {\rm LM6} &= &i \sigma^{(2)}_{ab} i \sigma^{(2)}_{cd} (\overline{Q}_a \lambda^A d_R) (\overline{Q}_{c}  \lambda^A d_R)
       ( \overline{\ell} _{b}  {\ell}^C_d)       \\
{\rm LM7 } &=&   (\overline{u}_R \gamma^\mu d_R)
	(\overline{u}_R \gamma_\mu d_R)
        (\overline{e}_R  e^C_R)  \\
{\rm LM8 } &=&  ( \overline{u}_R \gamma^\mu d_R) 
	 i \sigma^{(2)}_{ab} (\overline{Q}_{a}  d_R)
         ( \overline{\ell}_{b} \gamma_\mu  e^C_R)    \\ 
{\rm LM9 } &= &( \overline{u}_R \gamma^\mu \lambda^A d_R)
	 i \sigma^{(2)}_{ab} (\overline{Q}_{a}  \lambda^A d_R)
         ( \overline{\ell}_{b} \gamma_\mu  e^C_R) \\         
{\rm LM10} &=& (\overline{u}_R \gamma^\mu d_R) 
	(\overline{u}_R Q_a)
      (\overline{\ell}_a \gamma_\mu  e^C_R)  \\
 {\rm LM11} &= & (\overline{u}_R \gamma^\mu \lambda^A d_R) 
	(\overline{u}_R \lambda^A Q_a)
      (\overline{\ell}_a \gamma_\mu  e^C_R)      
    \label{eqn-basis-summary-dim9}
\eea 
\end{subequations}
The 11 operators above correspond to 11 of the 24 operators in Eqn.(\ref{4q2l-op}). At this leading order in $v/\Lambda$ only the following eleven 4-quark operators can appear, out of a possible set of 16: 
${\cal O}_{1LR}$, ${\cal O}^\lambda_{1LR}$, 
${\cal O}_{2RL}$, ${\cal O}^\lambda_{2RL}$,
${\cal O}_{2LR}$, ${\cal O}^\lambda_{2LR}$,
${\cal O}_{3R}$,  
${\cal O}^\mu_{RRLR}$, ${\cal O}^{\lambda \mu}_{RRLR}$, 
${\cal O}^\mu_{RRRL}$, and ${\cal O}^{\lambda \mu}_{RRRL}$. The results obtained here suggest that lattice QCD efforts to study the matrix elements of 4-quark operators relevant to a neutrinoless double $\beta$ decay signal should focus on this set of operators. We note that out of the 8 scalar operators that are allowed by $SU(3)_c \times U(1)_{em}$ invariance, 7 of these operators
are allowed,  by the full electroweak invariance of the theory,  at LO in $v/\Lambda$. To LO the only operator that does not appear is ${\cal O}_{3L}$.
The results of this Section are summarized in Table \ref{BigTable1}. 

\begin{table*}
\centering
\resizebox{\textwidth}{!}{
\begin{tabular}{c|c|c|c|c|c|c|}
\hline \hline
\multirow{2}{*}{operator} & \multirow{2}{*}{content} & \multicolumn{3}{c|}{hadron collider signatures} & \multirow{2}{*}{Low Energy}  & \multirow{2}{*}{$\chi$PT $(\pi \pi)$} \\ 
\cline{3-5}
& & \multicolumn{1}{p{2cm}|}{same-sign dilepton} &   \multicolumn{1}{c|}{$e$+MET} &  \multicolumn{1}{p{2.5cm}|}{dijet+ MET} & & \\
\hline 
\multicolumn{7}{c}{dimension 9} \\ 
\hline
 & & & & & & \\ 
LM1 & $ 
 i \sigma^{(2)}_{a b} (\overline{Q}_{a}  \gamma^\mu Q_{c})
     ( \overline{u}_R  \gamma_\mu d_R) 
     ( \overline{\ell}_b  \ell^C _c ) $ & $\surd$ &  $\surd$   &  $\surd$  & ${\cal O}_{1LR} \otimes (LL)$ & LO \\ 
     & & & & &  & \\
LM2 & $ 
 i \sigma^{(2)}_{a b} (\overline{Q}_{a}  \gamma^\mu \lambda^A Q_{c})
     ( \overline{u}_R  \gamma_\mu \lambda^A d_R) 
     ( \overline{\ell}_b  \ell^C _c ) $ & $\surd$ &  $\surd$   &  $\surd$  & ${\cal O}^\lambda_{1LR} \otimes (LL)$ & LO \\ 
     & & & & &  & \\     
LM3 & $      (\overline{u}_R Q_{a} ) (\overline{u}_R Q_{b})
     (   \overline{\ell}_{a}  \ell^C_{b}) $ & $\surd$ &  $\surd$   &  $\surd$   & ${\cal O}_{2RL} \otimes (LL)$ & LO \\ 
     & & & & & & \\
LM4 & $      (\overline{u}_R \lambda^A Q_{a} ) (\overline{u}_R \lambda^A Q_{b})
     (   \overline{\ell}_{a}  \ell^C_{b}) $ & $\surd$ &  $\surd$   &  $\surd$   & ${\cal O}^\lambda_{2RL} \otimes (LL)$ & LO \\ 
     & & & & & & \\     
LM5  & $ i \sigma^{(2)}_{ab} i \sigma^{(2)}_{cd} (\overline{Q}_a d_R) (\overline{Q}_{c} d_R)
       ( \overline{\ell} _{b}  {\ell}^C_d) $ & $\surd$  &  $\surd$   &  $\surd$  & ${\cal O}_{2LR} \otimes (LL)$   & LO \\ 
       & & & & & & \\
LM6  & $ i \sigma^{(2)}_{ab} i \sigma^{(2)}_{cd} (\overline{Q}_a \lambda^A d_R) (\overline{Q}_{c} \lambda^A d_R)
       ( \overline{\ell} _{b}  {\ell}^C_d) $ & $\surd$  &  $\surd$   &  $\surd$  & ${\cal O}^\lambda_{2LR} \otimes (LL)$   & LO \\ 
       & & & & & & \\       
LM7 & $    (\overline{u}_R \gamma^\mu d_R)
	(\overline{u}_R \gamma_\mu d_R)
        (\overline{e}_R  e^C_R) $ & $\surd$ &  $\ddot{\frown}$ &  $\ddot{\frown}$   & ${\cal O}_{3R} \otimes (RR)$  & NNLO  \\ 
        & & & & & & \\      
LM8 & $   ( \overline{u}_R \gamma^\mu d_R)
	 i \sigma^{(2)}_{ab} (\overline{Q}_{a}  d_R)
         ( \overline{\ell}_{b} \gamma_\mu e^C_R)  $ & $\surd$ & $\surd$ & $\ddot{\frown}$ & ${\cal O}^\mu_{RRLR}\otimes  (LR)$ & - \\ 
         & & & & & & \\               
LM9 & $   ( \overline{u}_R \gamma^\mu \lambda^A d_R)
	 i \sigma^{(2)}_{ab} (\overline{Q}_{a}  \lambda^A d_R)
         ( \overline{\ell}_{b} \gamma_\mu e^C_R)  $ & $\surd$ & $\surd$ & $\ddot{\frown}$ & ${\cal O}^{\lambda \mu}_{RRLR}\otimes  (LR)$ & - \\ 
         & & & & & & \\          
LM10  & $   (\overline{u}_R \gamma^\mu d_R) 
	(\overline{u}_R Q_a)
      (\overline{\ell}_a \gamma_\mu  e^C_R)   $ & $\surd$ & $\surd$ & $\ddot{\frown}$  & ${\cal O}^\mu_{RRRL}\otimes  (LR)$ & - \\     
         & & & & & & \\    
LM11  & $   (\overline{u}_R \gamma^\mu \lambda^A d_R) 
	(\overline{u}_R \lambda^A Q_a)
      (\overline{\ell}_a \gamma_\mu  e^C_R)   $ & $\surd$ & $\surd$ & $\ddot{\frown}$  & ${\cal O}^{\lambda \mu}_{RRRL}\otimes  (LR)$ & - \\     
         & & & & & & \\           
         & & & & & & \\    
 \hline \hline         
 \end{tabular}}
\caption{Table of dimension-9 electroweak invariant operators contributing to $0 \nu \beta \beta$ decay and hadron collider processes. A `$\surd$' indicates the operator contributes to the hadron collider process, whereas a `$\ddot{\frown}$' indicates that it does not. In the ``Low Energy" column the notation $LL$, $LR$, and $RR$ refer to whether the two leptons in the operator are $\overline{e}_L e^C_L$, $\overline{e}_L \gamma^\mu e^C_R$, or $\overline{e}_R e^C_R$, respectively. For a given operator , the last column indicates at what chiral order the two-pion interactions first appear, using the results summarized in Tables 
\ref{table:scalar} and \ref{table:vector}. A `-' indicates the operator does not contribute to NNLO order.
\label{BigTable1}}
\end{table*}

\section{Mapping onto chiral perturbation theory} 
\label{sec:chiralPT}

\begin{figure}
\centering
\includegraphics[width=0.8\textwidth]{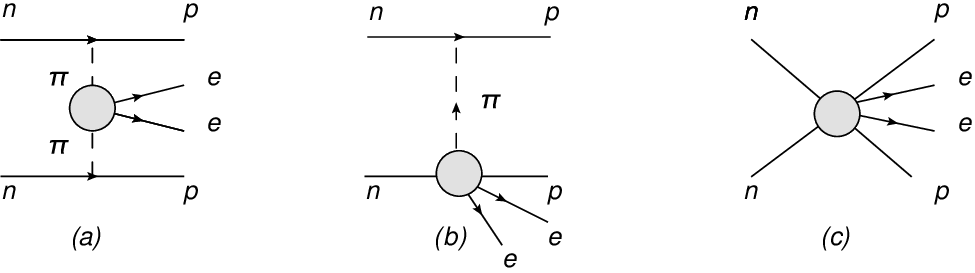}
\caption{Representative Feynman diagrams contributing to neutrinoless double beta decay inside of a nucleus. Diagram (a) due to induced 
$\pi \pi ee$ vertices, (b) due to induced $\pi NN ee$ vertices, and (c) due to induced $NNNNee$ vertices.}
\label{fig:chiPT}
\end{figure}

The next step is to obtain the effective Hamiltonian of these interactions inside a nucleus, using chiral perturbation theory ($\chi$PT) to match the effective theory at the GeV scale onto 
the effective theory involving pions and nucleons defined below that scale. The application of $\chi$PT to neutrinoless double $\beta$ was pioneered and developed in Ref. \cite{Prezeau:2003xn}. The processes relevant to neutrinoless double $\beta$ decay are shown in Figure \ref{fig:chiPT}. 
The strength of the 
contact interaction involving two electrons to pions and nucleons can only be determined accurately using lattice QCD. As noted in the Introduction, preliminary lattice results for the $\pi \pi$ matrix elements now exist \cite{Nicholson:2016byl}. Approximate chiral $SU(3)$ symmetry can also be used to estimate the same $\pi \pi$ matrix elements, by relating them 
to 
$K \pi \pi$ \cite{Savage:1998yh} \cite{Cirigliano:2017ymo} and
$K^0$-$\overline{K}^0$ \cite{Cirigliano:2017ymo} matrix elements, which have been computed using lattice QCD.


The $\chi$PT formalism organizes the effective theory into a simultaneous expansion in $\partial/(4 \pi f_\pi)$ and $m_\pi/(4 \pi f_\pi)$, where $\partial \sim m_\pi$ is a typical momentum transfer.
Since quarks couple to all hadrons, intuitively these 4-quark operators will induce couplings of the lepton bilinear to nucleons and importantly, to pions. In the power counting, the neutrinoless double beta decay rate will be formally dominated by the long-distance contribution caused by the exchange of pions, rather than the direct coupling of the two leptons to a 4-nucleon contact operator. 

This intuition can be formalized by the explicit power counting of the diagrams shown in Fig. \ref{fig:chiPT} that contribute to a $0 \nu \beta \beta$ decay signal inside of a nucleus. The power counting involves two parts. First, the quark operators ${\cal O}$ in Eqs. (\ref{scalarO1LR}-\ref{vectorOlRRRL}) are mapped onto all possible hadronic operators $\tilde{\cal O}$ that have the same transformation properties under $SU(2)_L \times SU(2)_R$, with each such operator $\tilde{\cal O}$ appearing at some given chiral order. The interactions $\tilde{\cal O}$ important to neutrinoless double $\beta$ decay are vertices containing two pions, a pion and two nucleons, or four nucleons. Then the operators $\tilde{\cal O} e e $ are inserted into a diagram involving four external nucleons and two electrons, as shown in Fig. \ref{fig:chiPT}. If we denote the chiral order of $\tilde{\cal O}$ as $n^L_{\tilde{\cal O}}$, where 
$n^L_{\tilde{\cal O}}$ can be found from Tables \ref{table:scalar} or \ref{table:vector}, then the chiral order of diagram (a) which has an insertion of a $\pi \pi e e$ interaction is $n^L_{\pi\pi}-2$, diagram (b) which has a $\pi N N e e$ interaction is $n^L_{\pi NN}-1$, and diagram (c) with 
an $ NN N N e e$ interaction insertion is $n^L_{N^4}$. An operator that contributes to a $\pi \pi ee$ interaction at lowest order is seen to be ``enhanced" in the neutrinoless double $\beta$ decay amplitude due to the long-distance pion exchange, as compared to its lowest order contribution to the other vertices. 

Inspecting the last column in Table 
\ref{BigTable1},
the power scaling of the amplitude due to an insertion of one of the operators LM1--LM6 is $p^{-2}$ and is dominated by the LO contribution those operators make to the $\pi \pi ee $ interaction.  The amplitude for an insertion of an operator in LM8--LM13  has a higher chiral order since none of these operators contribute at LO to the 
$\pi \pi ee$ interaction, as we show below. 

To map quark operators ${\cal O}$ onto $\tilde{\cal O}$ operators in the effective chiral theory, we follow Ref. \cite{Savage:1998yh} and write each 4-quark operator as 
\beq 
{\cal O} = T^{ab}_{cd} (\overline{q}^c \Gamma q_a) (\overline{q}^d \Gamma^\prime q_b) 
\eeq 
for some Dirac matrices $\Gamma$ and $\Gamma^\prime$. Here $a,b$ are $SU(2)$ flavor indices, 
\beq 
T^{ab}_{cd} = (\tau^+)^{~a}_c (\tau^+)^{~b}_d,
\eeq 
and $q_L=(u ~d)_L$ and $q_R=(u ~d)_R$. 
The transformation of $T$ under $SU(2)_L \times SU(2)_R$ is then determined by the transformations of the quarks, 
 \footnote{That is, $q_{La} \rightarrow L_a^b q_{Lb}$, $\overline{q}^a_L \rightarrow \overline{q}^c_{L} L^{\dagger a}_c$, where $L^{\dagger a}_c L^b _a = \delta^b_c$, $L^{\dagger a}_b \equiv (L^*)^b_a$, and similar relations for $L \rightarrow R$. }
and by the requirement that ${\cal O}$ is invariant under the chiral symmetry. This means that the transformation of $T$ will involve a product of $L$s and $R$s, 
\beq 
T \rightarrow T \otimes X_1 \otimes X_2    \otimes  X_3 \otimes X_4  
\eeq 
where each $X_i$ is $L^\dagger$ for a $q_L$,  $R^\dagger$ for a $q_R$, and $L$ for a $\overline{q}_L$,  $R$ for a $\overline{q}_R$. Four such $X$'s appear, one for each quark in the operator ${\cal O}$. Quark operators ${\cal O}$ that differ in their Dirac matrices $\Gamma$ and $\Gamma^\prime$ but that have the same $SU(2)_L \times SU(2)_R$ transformations will map onto the same chiral operators $\tilde{\cal O}$, and appear in the low-energy theory with different low-energy constants. 

In the chiral theory, one forms operators $\tilde{\cal O}(\pi, N)$ out of the pions ($\pi$) and nucleons ($N$) such that $T^{ab}_{cd} \tilde{\cal O}^{cd}_{ab}(\pi, N)$
is chirally invariant. 
In general a single 4-quark operator will map onto multiple operators in the chiral theory. 

To illustrate, 
recall that under $SU(2)_L \times SU(2)_R$, the pion composite field  $\xi = \hbox{Exp} [\pi \cdot \tau/2F_\pi]$ has a bilinear transformation $\xi \rightarrow L \xi U^\dagger=U \xi R^\dagger$, and the nucleon field $N$ transforms linearly, $N \rightarrow U N$, for unitary $L$, $R$ and $U(\xi,L,R)$. The $\tau^a$'s are the Pauli matrices. To each operator ${\cal O}$ we have the transformation of $T$ as described above. Using this $T$, we first construct a ``proto" $\tilde{\cal O}$ out of products of $\xi$ and $\xi^\dagger$'s, such that under $SU(2)_L \times SU(2)_R$, the explicit dependence of $L$ and $R$'s exactly cancels. 

One can also create additional $proto-\tilde{\cal O}$  objects at higher chiral order by inserting derivatives ${\cal D}_\mu$ or quark masses. 
Recall the derivative is ${\cal D}_\mu \equiv \partial_\mu +i {\cal V}_\mu$ with ${\cal V}_\mu=-i (\xi^\dagger \partial_\mu \xi + \xi \partial_\mu \xi^\dagger)/2$. Under chiral transformations ${\cal V}_\mu \rightarrow U({\cal V}_\mu + i \partial_\mu U)U^\dagger$, such that ${\cal D}_\mu \xi \rightarrow U ({\cal D}_\mu \xi) R^\dagger$ transforms the same as $\xi$.
Since the quark mass matrix transforms as $m_q \rightarrow R m_q L^\dagger$, the combination $m_q \xi$ transforms as $m_q \xi \rightarrow R(m_q \xi) U^\dagger$ which is the same transformation property as $\xi^\dagger$. This means that for any $proto-\tilde{\cal O}$ operator generated using the method described in the previous paragraph, new operators with higher chiral order can be created by substituting $\xi \rightarrow {\cal D}_\mu \xi$, $\xi^\dagger \rightarrow m_q \xi$ or $\xi \rightarrow \xi^\dagger m_q^\dagger$.

Because $\xi$ and $\xi^\dagger$ transform bilinearly, the ``proto"-$\tilde{\cal O}$ will not be invariant, but will instead transform like 
\bea 
(proto-\tilde{\cal O} )&\rightarrow & (proto-\tilde{\cal O}) \otimes Y_1 \otimes Y_2 \otimes Y_3 \otimes Y_4
\label{proto-O-transform}
\eea 
where each $Y_i$ is either a $U$ or $U^\dagger$. 

To obtain invariants, one simply does the following. 
\begin{itemize} 
\item To obtain $\tilde{O}$ involving only pion fields, contract the four ``free" indices of the $proto-\tilde{\cal O}$ in all possible ways so that the product of $U$'s and $U^\dagger$'s in 
Eqn. (\ref{proto-O-transform}) give the identity because $U$ is unitary. 
\item To obtain $\tilde{O}$ involving pion fields and only {\em two} nucleons, multiply two of the free indices of  $proto-\tilde{\cal O}$ by a nucleon and anti-nucleon, in all possible combinations, so that a $U$ and $U^\dagger$ in Eqn. (\ref{proto-O-transform}) cancel. Contract the remaining two free indices to form an invariant. 
This procedure will generate chirally invariant two nucleon operators. 
\item To obtain $\tilde{O}$ involving pion fields and {\em four} nucleons, multiply the free indices of  $proto-\tilde{\cal O}$ by two nucleons and two-anti nucleons in all inequivalent combinations, such that the $U$ and/or $U^\dagger$'s cancel under the $SU(2)_L \times SU(2)_R$ transformation. This procedure will generate 
chirally invariant four-nucleon operators. 
\item New operators having higher chiral order can also be generated. One can iterate the process described above by substituting derivatives, $\xi \rightarrow {\cal D}_\mu \xi$, or a quark mass matrix, $\xi^\dagger \rightarrow m_q \xi$ or $\xi \rightarrow \xi^\dagger m_q^\dagger$. Or one can also multiply in by other chirally invariant operators such as 
tr$({\cal D}^\mu \xi {\cal D}_\mu \xi^\dagger)$, etc. Finally, 
one can form operators out of additional products of $T$ or $T^\dagger$'s, but the net number of $T$'s $-T^\dagger$'s$=1$ to conserve electric charge. 
\end{itemize}

We focus for the rest of this section on operators $\tilde{\cal O}$ 
that only involve pion fields and no nucleons. The reason is that formally in the power counting these vertices give the dominant contribution to the $0 \nu \beta \beta$ decay amplitude in a nucleus, as discussed previously.  
We note that the $proto-\tilde{\cal O}$ obtained for each ${\cal O}_i$ in the sections that follow are still useful beyond the context of determining the $\pi \pi$ operators, for they are also needed as a first step in systematically determining the $\pi NN$ and $NNNN$ operators. 
We next consider each operator ${\cal O}_i$ in turn. In the subsections that follow we set $F_\pi=1$. 

\subsection{Scalar $\pi \pi$ operators} 

\subsubsection{${\cal O}_{1LR}$,  ${\cal O}^\lambda_{1LR}$} 

Since these three operators have the same $SU(2)_L \times SU(2)_R$ transformation properties, they each map onto the same chiral operators. Because of that we will only discuss one such operator, with the implication that our results apply to the other operator. We will be illustrative in this way throughout this subsection and the one that follows. 

With 
\begin{align*}
{\cal O}_{1LR} &\equiv (\overline{q}_L \gamma^\mu \tau^+ q_L) (\overline{q}_R \gamma_\mu \tau^+ q_R), \\
T^{ab}_{cd} &\rightarrow T ^{\alpha \beta}_{\rho \sigma} L ^\rho_c R ^\sigma_d  L^{\dagger a}_\alpha R^{\dagger b}_\beta \\
proto-\tilde{\cal O}_{1LR}&= T^{ab}_{cd} \xi^i_a \xi ^{\dagger j}_b \xi^{\dagger c}_k  \xi ^d _l 
\end{align*}
where to be specific $\xi^i _a \rightarrow L^b_a \xi^k_b U^{\dagger i}_k=U^b_a \xi^k_b R^{\dagger i}_k$, 
$\xi^{\dagger j}_b \rightarrow U_b^c \xi^{\dagger k}_c L^{\dagger j}_k=R_b^c \xi^{\dagger k}_c U^{\dagger j}_k$.
To obtain an invariant contract $i$ with $k$ and $j$ with $l$, or $i$ with $l$ and $k$ with $j$. The former option gives 
a vanishing double trace operator, $\hbox{tr}(\xi^\dagger \tau^+ \xi) \hbox{tr}(\xi \tau^+ \xi^\dagger)=0$ since $\xi \xi^\dagger=\mathbb{1}$. 
The latter option gives $\hbox{tr}(\xi \xi \tau^+ \xi^\dagger \xi^\dagger \tau^+)=2 \pi^- \pi^- + \cdots$.
Thus ${\cal O}_{1LR}$ gives one pion operator at LO. 

\subsubsection{${\cal O}_{2RL}$, ${\cal O}^\lambda_{2RL}$}
With 
\begin{align*}
{\cal O}_{2RL} &\equiv (\overline{q}_R  \tau^+ q_L) (\overline{q}_R  \tau^+ q_L), \\
T ^{ab}_{cd} &\rightarrow T ^{\alpha \beta}_{\rho \sigma} R ^\rho_c R ^\sigma_d  L^{\dagger a}_\alpha L^{\dagger b}_\beta \\ 
proto-\tilde{\cal O}_{2RL}&= T^{ab}_{cd} \xi^i_a \xi ^{j}_b \xi^{c}_k  \xi ^d _l 
\end{align*}
As before, the two possible contractions give $\hbox{tr}(\tau^+ \xi \xi \tau^+ \xi \xi)=-2 \pi^- \pi^- +O(\pi^3)$, and a double 
trace operator $\hbox{tr}(\tau^+ \xi \xi) \hbox{tr}(\tau^+ \xi \xi) =(\sqrt{2} i \pi^-)^2+O(\pi^3)$. 
Thus ${\cal O}_{2RL}$ gives two pion operators at LO.

\subsubsection{${\cal O}_{2LR}$, ${\cal O}^\lambda_{2LR}$}
With 
\begin{align*}
{\cal O}_{2LR} & \equiv (\overline{q}_L  \tau^+ q_R) (\overline{q}_L  \tau^+ q_R)\\ 
T ^{ab}_{cd} &\rightarrow T ^{\alpha \beta}_{\rho \sigma} L ^\rho_c L ^\sigma_d  R^{\dagger a}_\alpha R^{\dagger b}_\beta \\
proto-\tilde{\cal O}_{2LR} &= T^{ab}_{cd} \xi^{i \dagger}_a \xi ^{\dagger j}_b \xi^{\dagger c}_k  \xi ^{\dagger d} _l 
\end{align*}
Likewise, here there are two invariant operators, 
 $\hbox{tr}(\tau^+ \xi^\dagger \xi^\dagger \tau^+ \xi^\dagger \xi^\dagger)=-2 \pi^- \pi^- +O(\pi^3)$, and a double 
trace operator $\hbox{tr}(\tau^+ \xi^\dagger \xi^\dagger) \hbox{tr}(\tau^+ \xi^\dagger \xi^\dagger)=(-\sqrt{2} i \pi^-)^2+O(\pi^3) $
Thus ${\cal O}_{2LR}$ gives two pion operators at LO.

\subsubsection{${\cal O}_{3L}$} 
With 
\begin{align*}
{\cal O}_{3L} &\equiv (\overline{q}_L \gamma^\mu \tau^+ q_L) (\overline{q}_L\gamma_\mu \tau^+ q_L)\\
T^{ab}_{cd} &\rightarrow T^{\alpha \beta}_{\rho \sigma} L^\rho_c L^\sigma _d L^{\dagger a} _\alpha L^{\dagger b}_\beta  
\end{align*}
so two $\xi$'s and two $\xi^\dagger$'s are needed, 
\beq 
proto-\tilde{\cal O} _{3L}= T^{ab}_{cd} \xi^i_a \xi^j _b \xi^{\dagger c} _k \xi ^{\dagger d}_l 
\eeq 
To construct an invariant, we either can contract $i$ with $k$ and $j$ with $l$, or $i$ with $l$ and $j$ with $k$. With 
$T^{ab}_{cd} = (\tau^{+})^a_c (\tau^+)^b_d$, the first possibility gives a double trace operator 
$\hbox{tr}(\xi^\dagger \tau^+ \xi) \hbox{tr}(\xi^\dagger \tau^+ \xi)$ which vanishes, because $\xi \xi^\dagger=\mathbb{1}$.  The second possibility 
gives $\hbox{tr} (\xi^\dagger \tau^+ \xi \xi ^\dagger \tau^+ \xi)$. However, for the same reasons this operator also vanishes. 

{\em Therefore to leading chiral order (LO), ${\cal O}_{3L}$ does not have any purely pionic operators.} This finding confirms in a more systematic manner the same conclusion reached by Ref. \cite{Prezeau:2003xn}. Turning to NNLO, here one does find non-vanishing operators obtained by applying derivatives. One has three single trace operators,
\beq 
\hbox{tr}({\cal D}^\mu \xi^\dagger \tau^+ {\cal D}_\mu \xi \xi^\dagger \tau^+ \xi), 
~\hbox{tr}({\cal D}^\mu \xi^\dagger \tau^+ \xi {\cal D}_\mu \xi^\dagger \tau^+ \xi), 
~\hbox{tr}(\xi^\dagger \tau^+ {\cal D}^\mu \xi \xi^\dagger \tau^+ {\cal D}_\mu \xi), 
\label{3L-NNLO-str}
\eeq 
and three double trace operators 
 \beq 
\hbox{tr}({\cal D}^\mu \xi^\dagger \tau^+ \xi) \hbox{tr}({\cal D}_\mu \xi^\dagger \tau^+ \xi), \hbox{tr}(\xi^\dagger \tau^+ {\cal D}^\mu \xi) \hbox{tr}({\cal D}_\mu \xi^\dagger \tau^+ \xi), \hbox{tr}(\xi^\dagger \tau^+{\cal D}^\mu \xi) \hbox{tr}(\xi^\dagger \tau^+ {\cal D}_\mu \xi).
\label{3L-NNLO-dtr}
\eeq 
For instance, the first single trace operator in Eq. (\ref{3L-NNLO-str}) is $=(\partial^\mu \pi^-)^2/2+\cdots$ and the first double trace operator in 
Eq. (\ref{3L-NNLO-dtr}) is $=-(\partial^\mu \pi^-)^2/2+\cdots$.
To get a non-vanishing operator from inserting quark masses one needs to insert two quark mass matrices, because at LO both the single and double trace operators above have two instances of $\xi \xi^\dagger=\mathbb{1}$. Such operators are however beyond NNLO. 

\subsubsection{${\cal O}_{3R}$}
With 
\beq
{\cal O}_{3R} \equiv (\overline{q}_R \gamma^\mu \tau^+ q_R) (\overline{q}_R\gamma_\mu \tau^+ q_R)
\eeq 
this operator will have the same transformation properties as ${\cal O}_{3L}$ except with $L \rightarrow R$. Like 
${\cal O}_{3L}$, it will not have any purely pionic operators at LO. More explicitly, 
\beq 
T^{ab}_{cd} \rightarrow T^{\alpha \beta}_{\rho \sigma} R^\rho_c R^\sigma _d R^{\dagger a} _\alpha R^{\dagger b}_\beta  
\eeq 
and since under $L \rightarrow R$, $\xi \rightarrow \xi^\dagger$,
\beq 
proto-\tilde{\cal O}_{3R} = T^{ab}_{cd} \xi^{\dagger i}_a \xi^{\dagger j} _b \xi^{c} _k \xi ^{d}_l 
\eeq 
The two possible contractions in this case give $\hbox{tr}(\xi \tau^+ \xi^\dagger) \hbox{tr}(\xi \tau^+ \xi^\dagger)=0$
and $\hbox{tr} (\xi \tau^+ \xi^\dagger \xi \tau^+ \xi^\dagger)=0$. 

{\em ${\cal O}_{3R}$ does not have any purely pionic operators to LO.} This result is not surprising given the previous result for ${\cal O}_{3L}$ and the parity invariance of the QCD interactions. 
At NNLO one finds 3 double trace and 3 single trace operators, and no operators involving quark masses, just like with ${\cal O}_{3L}$. These operators can be obtained from Eqs. (\ref{3L-NNLO-str}) and 
(\ref{3L-NNLO-dtr}) by the substitution $\xi \longleftrightarrow \xi^\dagger$. 

The results for the scalar operators are summarized in Table \ref{table:scalar}.

\begin{table*} 
\centering
{
\begin{tabular}{c|c|c|}
\hline \hline
 & \multicolumn{2}{c}{scalar 4-quark operator} \\
 \hline
  &   ${\cal O}_{1LR}$, ${\cal O}^\lambda_{1LR}$, ${\cal O}_{2LR}$,  ${\cal O}^\lambda_{2LR}$, ${\cal O}_{2RL}$, ${\cal O}^\lambda_{2RL}$  & ${\cal O}_{3L}$, ${\cal O}_{3R}$ \\
  \hline
 $n^{\cal O}_{\pi \pi}$ & LO & NNLO \\
 \hline \hline 
 \end{tabular}}
 \caption{Chiral order of $\pi \pi$ interactions induced by the 8 scalar operators ${\cal O}$. \label{table:scalar}}
 \end{table*}

\subsection{Vector $\pi \pi$ operators} 

It was noted in \cite{Prezeau:2003xn} that 
these operators don't contribute at LO. The reason is that at LO the effective Lagrangian must be of the form
\beq 
\pi^- \partial^\mu \pi^- \overline{e}  \gamma^\mu \gamma_5 e^c ~.
\label{vector:LOpipi}
\eeq 
By an integration of parts in the effective Lagrangian and use of the electron equations of motion, this operator is seen to be proportional to electron mass and
can be neglected.

At NNLO one can show that the same manipulations can be used to make all the two-pion vector operators proportional to the electron mass. 
For instance, one has operators such as 
\beq 
\partial^2 \pi^- \partial^\mu \pi^- \overline{e} \gamma^\mu  \gamma_5 e^c ~,  \partial^\nu\pi^- \partial _\nu \partial^\mu \pi^- \overline{e}  \gamma^\mu \gamma_5 e^c
\eeq 
but here one can use the pion equations of motion \cite{Politzer:1980me} and integration by parts to again obtain an operator of the form that appears in Eq. (\ref{vector:LOpipi}), which can be neglected.  

In the following subsections we identify the purely pionic $\tilde{\cal O}^\mu$ operators through to NNLO. The reason for doing this is that as an intermediate step the proto-$\tilde{O}^\mu$ operators are constructed, which can be used as a basis for determining the determining the 
inequivalent two nucleon and four nucleon operators. 

\subsubsection{${\cal O}^\mu_{LLLR}$, ${\cal O}^{\lambda \mu}_{LLLR}$}
With 
\begin{subequations}
\beq 
{\cal O}^\mu_{LLLR} \equiv (\overline{q}_L \gamma^\mu \tau^+ q_L)(\overline{q}_L \tau^+ q_R) 
\eeq 
\beq 
T^{ab}_{cd} \rightarrow T^{\alpha \beta}_{\rho \sigma} L^\rho_c L^\sigma _d L^{\dagger a} _\alpha R^{\dagger b}_\beta  
\eeq 
\beq 
proto-\tilde{\cal O}^\mu _{LLLR} = {\cal D}^\mu \otimes T^{ab}_{cd}  \xi^{i}_a \xi^{\dagger j} _b \xi^{\dagger c} _k \xi ^{\dagger d}_l
\label{LLLR-proto-O}
\eeq
\end{subequations}
where ${\cal D}^\mu \otimes$ means the four independent operators formed out of ${\cal D}^\mu $ contracted with a  $\xi$ or a $\xi^\dagger$. By the linearity of ${\cal D}$, operators such as ${\cal D} (\xi \xi), {\cal D} (\xi \xi \xi)$, etc., are not independent from operators having the derivative act on a single $\xi$. Because ${\cal D}_\mu \xi$ has the same transformation properties as $\xi$, 
we can construct invariants in the same way as before, leading to a set of single and double trace operators, each involving a single derivative at LO. 
One can obtain proto-operators at NNLO chiral order by either applying two derivatives ${\cal D}^\nu \otimes {\cal D}_\nu$ in all possible ways, or insert a single quark mass matrix, to 
the proto-operators in Eq. (\ref{LLLR-proto-O}).

\subsubsection{${\cal O}^\mu_{RRLR}$, ${\cal O}^{\lambda \mu}_{RRLR}$}

One finds  
\begin{subequations}
\beq 
{\cal O}^\mu_{RRLR} \equiv (\overline{q}_R \gamma^\mu \tau^+ q_R)(\overline{q}_L \tau^+ q_R) 
\eeq 
\beq 
T^{ab}_{cd} \rightarrow T^{\alpha \beta}_{\rho \sigma} R^\rho_c L^\sigma _d R^{\dagger a} _\alpha R^{\dagger b}_\beta  
\eeq 
\beq 
proto-\tilde{\cal O}^\mu_{RRLR} = {\cal D}^\mu \otimes T^{ab}_{cd}  \xi^{\dagger i}_a \xi^{\dagger j} _b \xi^{c} _k \xi ^{\dagger d}_l
\eeq
\end{subequations}



\subsubsection{${\cal O}^\mu_{LLRL}$, ${\cal O}^{\lambda \mu}_{LLRL}$}
One finds 
\begin{subequations}
\beq 
{\cal O}^\mu_{LLRL} \equiv (\overline{q}_L \gamma^\mu \tau^+ q_L)(\overline{q}_R \tau^+ q_L) 
\eeq 
\beq 
T^{ab}_{cd} \rightarrow T^{\alpha \beta}_{\rho \sigma} L^\rho_c R^\sigma _d L^{\dagger a} _\alpha L^{\dagger b}_\beta  
\eeq 
\beq 
proto-\tilde{\cal O}^\mu_{LLRL} = {\cal D}^\mu \otimes T^{ab}_{cd}  \xi^{i}_a \xi^{j} _b \xi^{\dagger c} _k \xi ^{d}_l
\eeq
\end{subequations}


\subsubsection{${\cal O}^\mu_{RRRL}$, ${\cal O}^{\lambda \mu}_{RRRL}$}
One finds 
\begin{subequations}  
 \beq 
{\cal O}^\mu_{RRRL} \equiv (\overline{q}_R \gamma^\mu \tau^+ q_R)(\overline{q}_R \tau^+ q_L) 
\eeq 
\beq 
T^{ab}_{cd} \rightarrow T^{\alpha \beta}_{\rho \sigma} R^\rho_c R^\sigma _d R^{\dagger a} _\alpha L^{\dagger b}_\beta  
\eeq 
\beq 
proto-\tilde{\cal O}^\mu_{RRRL} = {\cal D}^\mu \otimes T^{ab}_{cd}  \xi^{\dagger i}_a \xi^{j} _b \xi^{c} _k \xi ^{d}_l
\eeq
\end{subequations}


These results for the vector operators are summarized in Table \ref{table:vector}.

\begin{table*} 
\centering
{
\begin{tabular}{c|c}
\hline \hline
 & (vector 4-quark operator) $\otimes~ \overline{e} \gamma_\mu \gamma_5 e^c$ \\
 \hline
  &   ${\cal  \tilde O}$ , ${\cal \tilde O}^{\lambda}_* $ \\ 
  \hline
 $n^{\cal O}_{\pi \pi}$ &  - \\
 \hline \hline 
 \end{tabular}}
 \caption{Chiral order of $\pi \pi e e$ interactions induced by vector operators ${\cal \tilde O}_* \equiv {\cal O}^\mu_* ~\overline{e} \gamma_\mu \gamma_5 e^c$ 
 and ${\cal \tilde O}^\lambda_* \equiv {\cal O}^{\lambda \mu}_* ~\overline{e} \gamma_\mu \gamma_5 e^c$, where $*=LLLR$, $RRLR$, $LLRL$ or $RRRL$. These interactions with two leptons can be eliminated to NNLO order as discussed in Ref. \cite{Prezeau:2003xn} and the text.\label{table:vector}}
 \end{table*}

\section{Dimension-11} 
\label{sec:Dimension-11}

The intent in this Section is find dimension-11 operators that after electroweak symmetry reduce to those operators in Eqn. (\ref{4q2l-op}) that have not previously been found at dimension-9. No attempt is made in this Section to find a complete basis at dimension-11, as for example one can always take a previously appearing dimension-9 operator and multiply it by $H^\dagger H$ to get a dimension-11 operator.  There are also qualitatively new operators that appear, such as the following electroweak invariant operator
\beq 
(\overline{u}_R \gamma^\mu T^a d_R) (\overline{u}_R \gamma^\nu d_R) G^a_{\mu \nu} \overline{e}_R e^C_R 
\eeq 

Twelve of the operators that don't appear at dimension-9 are made $SU(2)_L \times U(1)_Y$ gauge invariant through the insertion of additional Higgs fields, as follows.

\subsection{$\overline{e}_R e^C_R$}
   
\begin{align*}
{\rm LM12} &= i \sigma^{(2)}_{ab} H^*_{a} H^*_{c}(\overline{Q}_{b} \gamma^\mu Q_{c})
     (\overline{u}_R \gamma_\mu d_R)
     (  \overline{e}_R e^C_R)\\
     &\rightarrow - \frac{1}{2} v^2 (\overline{u}_L \gamma^\mu d_L)(\overline{u}_R \gamma_\mu d_R)(\overline{e}_R e^C_R) \\
 {\rm LM13} &= i \sigma^{(2)}_{ab} H^*_{a} H^*_{c}(\overline{Q}_{b} \gamma^\mu \lambda^A Q_{c})
     (\overline{u}_R \gamma_\mu \lambda^A d_R)
     (  \overline{e}_R e^C_R)\\     
{\rm LM14} &=  H^*_a H^*_b (\overline{u}_R Q_a) (\overline{u}_R Q_b)
       ( \overline{e}_R  e^C_R) \\
       & \rightarrow   \frac{1}{2} v^2 (\overline{u}_R d_L)(\overline{u}_R d_L) (\overline{e}_R e^C_R) \\
{\rm LM15} &=  H^*_a H^*_b (\overline{u}_R \lambda^A Q_a) (\overline{u}_R \lambda^A Q_b)
       ( \overline{e}_R  e^C_R) \\       
{\rm LM16} &=    i \sigma^{(2)}_{ab} i \sigma^{(2)}_{cd}   H^*_a H^*_c(\overline{Q}_b d_R)( \overline{Q}_d d_R) 
     (\overline{e}_R   e^C_R ) \\
     &\rightarrow   \frac{1}{2} v^2 (\overline{u}_L d_R)(\overline{u}_L d_R) (\overline{e}_R e^C_R) \\
 {\rm LM17} &=    i \sigma^{(2)}_{ab} i \sigma^{(2)}_{cd}   H^*_a H^*_c(\overline{Q}_b \lambda^A d_R)( \overline{Q}_d \lambda^A d_R) 
     (\overline{e}_R   e^C_R )     
\end{align*}
Above and in what follows, `$\rightarrow$' means `insert Higgs vev'. 
\subsection{$\overline{\ell} \ell^C$}
\begin{align*}
    { \rm LM18} &=   H_a H_b (\overline{u}_R \gamma^\mu d_R) 
	( \overline{u}_R \gamma_\mu d_R)
       ( \overline{\ell}_a   \ell^C_b) \\
       & \rightarrow   \frac{1}{2} v^2( \overline{u}_R \gamma^\mu d_R)( \overline{u}_R \gamma_\mu d_R) (\overline{e}_L e^C_L) 
\end{align*}

The next set of four operators contribute at low energy to the same $0 \nu \beta \beta$ operator (hence the same label), but contribute differently to lepton+MET and $2j+$ MET.     
\begin{align*}
{\rm LM19a} &= i  \sigma^{(2)}_{ae} i \sigma^{(2)}_{cf}  H^*_b H^*_d (\overline{Q}_a \gamma^\mu Q_b)  
       (\overline{Q}_c \gamma_\mu Q_d )  
       (\overline{\ell}_e  \ell^C_f) \\
       & \rightarrow   \frac{1}{2} v^2 \Big[ (\overline{u}_L \gamma^\mu d_L) (\overline{u}_L \gamma_\mu d_L) (\overline{e}_L e^C_L) - 2(\overline{u}_L \gamma^\mu d_L)(\overline{d}_L \gamma_\mu d_L) (\overline{e}_L \nu^C_L) \nonumber  \\
       &  ~~~~~~+ (\overline{d}_L \gamma^\mu d_L)(\overline{d}_L \gamma_\mu d_L)( \overline{\nu}_L \nu^C_L) \Big] ,   \\
{\rm LM19b} &=  i  \sigma^{(2)}_{ae} i\sigma^{(2)}_{cf}  H^*_e H^*_{d}(\overline{Q}_a \gamma^\mu Q_b)
        (\overline{Q}_{c} \gamma_\mu Q_{d} )
      ( \overline{\ell}_b  \ell^C_{f}) \\
      & \rightarrow   \frac{1}{2} v^2 \Big[(\overline{u}_L \gamma^\mu d_L) (\overline{u}_L \gamma_\mu d_L) (\overline{e}_L e^C_L) +(\overline{u}_L \gamma^\mu u_L - \overline{d}_L \gamma_\mu d_L)(\overline{u}_L \gamma_\mu d_L)(\overline{e}_L \nu^C_L) \nonumber \\
      &  ~~~~~~ - (\overline{u}_L \gamma^\mu u_L)(\overline{d}_L \gamma_\mu d_L)(\overline{\nu}_L \nu^C_L) \Big] , \\
{\rm LM19c} &=  i  \sigma^{(2)}_{ae } i\sigma^{(2)}_{cf} H^*_{e} H^*_{c} (\overline{Q}_{a} \gamma^\mu Q_{b})
        (\overline{Q}_{f} \gamma_\mu Q_{d} )
      (\overline{\ell}_{b}  \ell^C_{d} )\\
      & \rightarrow   - \frac{1}{2} v^2 \Big[ (\overline{u}_L \gamma^\mu d_L) (\overline{u}_L \gamma_\mu d_L) (\overline{e}_L e^C_L) +2 (\overline{u}_L \gamma^\mu u_L)(\overline{u}_L \gamma_\mu d_L)(\overline{e}_L \nu^C_L) 
      \nonumber \\
      &  ~~~~~~~~~~ + (\overline{u} \gamma^\mu u_L)(\overline{u}_L \gamma_\mu u_L)( \overline{\nu}_L \nu^C) \Big], \\
{\rm LM19d} &=   i\sigma^{(2)}_{ae} i\sigma^{(2)}_{cf} H^*_{c} H^*_{d} 
 (\overline{Q}_{e} \gamma^\mu  Q_{b})
        (\overline{Q}_{f} \gamma_\mu Q_{d})
       (\overline{\ell}_ {a}  \ell^C_{b}) \\
       & \rightarrow   \frac{1}{2} v^2 (\overline{u}_L \gamma^\mu d_L) \Big[(\overline{u}_L \gamma_\mu d_L) (\overline{e}_L e^C_L) +(\overline{u}_L \gamma_\mu u_L -\overline{d}_L \gamma_\mu d_L)(\overline{e}_L \nu^C_L), \nonumber \\ 
       &  ~~~~~~~~ - (\overline{d}_L \gamma_\mu u_L)(\overline{\nu}_L \nu^C_L) \Big] ~.
\end{align*}
The color-octet versions of the above operators are not independent, as shown in Section \ref{Appendix:complete-set}. 

\subsection{$\overline{\ell} \gamma^\mu e^C_R$}
The next three operators contribute to both SS dilepton and lepton+ MET, 
\bea 
{\rm LM20} &=&  i \sigma^{(2)}_{ae} i\sigma^{(2)}_{cd} H^*_{e} H^*_{b}  
	(\overline{Q}_{a} \gamma^\mu Q_{b})
        (\overline{Q}_{c} d_R)
        (\overline{\ell}_{d} \gamma_\mu  e^C_R) \\
        & \rightarrow &  \frac{1}{2} v^2 (\overline{u}_L \gamma^\mu d_L) \Big[ (\overline{u}_L d_R) (\overline{e}_L \gamma_\mu e^C_L) - (\overline{d}_L d_R)(\overline{\nu}_L \gamma_\mu e^C_R) \Big] \\
{\rm LM21} &=&  i \sigma^{(2)}_{ae} i\sigma^{(2)}_{cd} H^*_{e} H^*_{b}  
	(\overline{Q}_{a} \gamma^\mu \lambda^A Q_{b})
        (\overline{Q}_{c} \lambda^A d_R)
        (\overline{\ell}_{d} \gamma_\mu  e^C_R)        
 \eea
 The next  three operators contribute at low - energy to the same $0 \nu \beta \beta $ operator, but as with LM19(a-d), contribute differently to the other processes (in this case, only to lepton + MET). 
 \begin{align*}
{\rm LM22a} &=   i\sigma^{(2)}_{ab}  H^*_b H^*_{c}  
	(\overline{Q}_{a} \gamma^\mu Q_{c}) 
        ( \overline{u}_R Q_{d})
        (\overline{\ell}_{d}  \gamma_\mu  e^C_R )  \\
        & \rightarrow   \frac{1}{2} v^2 (\overline{u}_L \gamma^\mu d_L) \Big[ (\overline{u}_R d_L)(\overline{e}_L \gamma_\mu  e^C_R ) + (\overline{u}_R u_L)( \overline{\nu}_L \gamma_\mu e^C_R) \Big] \\
{\rm LM22b} &= i \sigma^{(2)}_{ab} H^*_{b} H^*_{d}  
	( \overline{Q}_{a} \gamma^\mu  Q_{c})
        (\overline{u}_R Q_{d}) 
        ( \overline{\ell} _{c} \gamma_\mu  e^C_R) \\
        & \rightarrow  \frac{1}{2} v^2 (\overline{u}_R d_L) \Big[ (\overline{u}_L \gamma^\mu d_L)(\overline{e}_L \gamma_\mu e^C_R) + (\overline{u}_L \gamma^\mu u_L) (\overline{\nu}_L \gamma_\mu e^C_R) \Big] \\
{\rm LM22c} &=  i \sigma^{(2)}_{ab} H^*_{c} H^*_{d}  
	( \overline{Q}_{a} \gamma^\mu  Q_{c})
        (\overline{u}_R Q_{d}) 
        ( \overline{\ell} _{b} \gamma_\mu  e^C_R) \\
        & \rightarrow  \frac{1}{2} v^2 (\overline{u}_R d_L) \Big[ (\overline{u}_L \gamma^\mu d_L)(\overline{e}_L \gamma_\mu e^C_R) - (\overline{d}_L \gamma_\mu d_L)(\overline{\nu}_L \gamma_\mu e^C_R) \Big]
\end{align*} 
We also have the color-octet versions 
 \begin{align*}
{\rm LM23a} &=   i\sigma^{(2)}_{ab}  H^*_b H^*_{c}  
	(\overline{Q}_{a} \gamma^\mu \lambda^A Q_{c}) 
        ( \overline{u}_R \lambda^A Q_{d})
        (\overline{\ell}_{d}  \gamma_\mu  e^C_R )  \\
{\rm LM23b} &= i \sigma^{(2)}_{ab} H^*_{b} H^*_{d}  
	( \overline{Q}_{a} \gamma^\mu  \lambda^A Q_{c})
        (\overline{u}_R \lambda^A Q_{d}) 
        ( \overline{\ell} _{c} \gamma_\mu  e^C_R) \\
{\rm LM23c} &=  i \sigma^{(2)}_{ab} H^*_{c} H^*_{d}  
	( \overline{Q}_{a} \gamma^\mu  \lambda^A Q_{c})
        (\overline{u}_R \lambda^A Q_{d}) 
        ( \overline{\ell} _{b} \gamma_\mu  e^C_R) ~.
\end{align*} 

\begin{table*}
\centering
\resizebox{\textwidth}{!}{
\begin{tabular}{c|c|c|c|c|c|c|}
\hline \hline
\multirow{2}{*}{operator} & \multirow{2}{*}{content} & \multicolumn{3}{c|}{hadron collider signatures} & \multirow{2}{*}{Low Energy} & \multirow{2}{*}{$\chi$PT $(\pi \pi)$} \\ 
\cline{3-5}
& & \multicolumn{1}{p{2cm}|}{same-sign dilepton} &   \multicolumn{1}{c|}{$e$+MET} &  \multicolumn{1}{p{2.5cm}|}{dijet+ MET} &  & \\
\hline 
\multicolumn{7}{c}{dimension 11} \\ 
\hline
 & & & & & & \\
LM12 & $  i \sigma^{(2)}_{ab} H^*_{a} H^*_{c}(\overline{Q}_{b} \gamma^\mu Q_{c})
     (\overline{u}_R \gamma_\mu d_R)
     (  \overline{e}_R  e^C_R) $ & $\surd$ &  $\ddot{\frown}$ &  $\ddot{\frown}$   & ${\cal O}_{1LR} \otimes  (RR)$  & LO \\ 
     & & & & & & \\
 LM13 & $  i \sigma^{(2)}_{ab} H^*_{a} H^*_{c}(\overline{Q}_{b} \gamma^\mu \lambda^A Q_{c})
     (\overline{u}_R \gamma_\mu \lambda^A d_R)
     (  \overline{e}_R  e^C_R) $ & $\surd$ &  $\ddot{\frown}$ &  $\ddot{\frown}$   & ${\cal O}^\lambda_{1LR} \otimes  (RR)$  & LO \\ 
     & & & & & & \\    
LM14 & $   H^*_a H^*_b (\overline{u}_R Q_a) (\overline{u}_R Q_b)
       ( \overline{e}_R  e^C_R) $ & $\surd$ &  $\ddot{\frown}$ &  $\ddot{\frown}$  & ${\cal O}_{2RL} \otimes (RR)$ & LO \\ 
       & & & & & & \\
LM15 & $   H^*_a H^*_b (\overline{u}_R \lambda^A Q_a) (\overline{u}_R \lambda^A Q_b)
       ( \overline{e}_R  e^C_R) $ & $\surd$ &  $\ddot{\frown}$ &  $\ddot{\frown}$  & ${\cal O}^\lambda_{2RL} \otimes (RR)$ & LO \\ 
       & & & & & & \\   
LM16 & $     i \sigma^{(2)}_{ab} i \sigma^{(2)}_{cd}   H^*_a H^*_c(\overline{Q}_b d_R)( \overline{Q}_d d_R) 
     (\overline{e}_R   e^C_R ) $ & $\surd$ &  $\ddot{\frown}$ &  $\ddot{\frown}$   &  ${\cal O}_{2LR} \otimes (RR)$ & LO \\  
     & & & & & & \\
LM17 & $     i \sigma^{(2)}_{ab} i \sigma^{(2)}_{cd}   H^*_a H^*_c(\overline{Q}_b \lambda^A d_R)( \overline{Q}_d  \lambda^A d_R) 
     (\overline{e}_R   e^C_R ) $ & $\surd$ &  $\ddot{\frown}$ &  $\ddot{\frown}$   &  ${\cal O}^\lambda _{2LR} \otimes (RR)$ & LO \\  
     & & & & & & \\
LM18  & $  H_a H_b (\overline{u}_R \gamma^\mu d_R) 
	( \overline{u}_R \gamma_\mu d_R)
       ( \overline{\ell}_a   \ell^C_b) $ & $\surd$ &  $\ddot{\frown}$  &  $\ddot{\frown}$  & ${\cal O}_{3R}\otimes (LL)$   & NNLO \\ 
       & & & & & & \\             
LM19a & $   i \sigma^{(2)}_{ae} i \sigma^{(2)}_{cf}  H^*_b H^*_d (\overline{Q}_a \gamma^\mu Q_b)  
       (\overline{Q}_c \gamma_\mu Q_d )  
       (\overline{\ell}_e  \ell^C_f) $ & $\surd$ &  $\surd$   &  $\surd$  &  ${\cal O}_{3L}\otimes (LL)$ & NNLO \\ 
       & & & & & & \\
LM19b & $ i  \sigma^{(2)}_{ae}i \sigma^{(2)}_{cf}  H^*_e H^*_{d}(\overline{Q}_a \gamma^\mu Q_b)
        (\overline{Q}_{c} \gamma_\mu Q_{d} )
      ( \overline{\ell}_b  \ell^C_{f}) $ & $\surd$ &  $\surd$   &  $\surd$  & ``\hbox{same as LM19a}"  & ``\hbox{same as LM19a}" \\  
      & & & & & & \\
LM19c & $    i \sigma^{(2)}_{ae } i\sigma^{(2)}_{cf} H^*_{e} H^*_{c} (\overline{Q}_{a} \gamma^\mu Q_{b})
        (\overline{Q}_{f} \gamma_\mu Q_{d} )
      (\overline{\ell}_{b}  \ell^C_{d} ) $ & $\surd$ &  $\surd$   &  $\surd$  & ``\hbox{same as LM19a}"  & ``\hbox{same as LM19a}" \\ 
      & & & & & &\\
LM19d  & $   i \sigma^{(2)}_{ae}i \sigma^{(2)}_{cf} H^*_{c} H^*_{d} 
 (\overline{Q}_{e} \gamma^\mu  Q_{b})
        (\overline{Q}_{f} \gamma_\mu Q_{d})
       (\overline{\ell}_ {a}  \ell^C_{b}) $ & $\surd$ &  $\surd$   &  $\surd$   & ``\hbox{same as LM19a}"   & ``\hbox{same as LM19a}"  \\ 
       & & & &  & & \\
 \hline
 \hline 
 \end{tabular}}
\caption{Table of dimension-11 electroweak invariant operators contributing to $0 \nu \beta \beta$ decay and hadron collider processes. 
 After restricting the Higgs field to its vev, these operators do not reproduce any of the operators appearing in Table \ref{BigTable1}. 
Same notation as Table \ref{BigTable1}. All the quark operators appearing in this Table are scalar.
\label{BigTable2}}
\end{table*}

In total there are 19 dimension-11 operators suppressed by an additional power of $v^2 /\Lambda^2$, that at low-energy lead to 12 of the operators in Eq. (\ref{4q2l-op}) that don't appear at dimension-9. The results of this Section are summarized in Tables \ref{BigTable2} and \ref{BigTable3}. 

\section{Dimension-13}
\label{sec:Dimension-13}
At dimension-9 we found 11 operators and at dimension-11 we found 12 more operators out of the complete set of 24 operators appearing in Eqns. (\ref{4q2l-op}).
That leaves one operator missing, namely $\sim (\overline{u}_L \gamma^\mu d_L) (\overline{u}_L \gamma^\mu d_L)  (\overline{e}_R e^C_R)$ and it only involves left-handed quarks $Q_L$. Since the lepton bilinear has hyper-charge 2 and the part of the operator involving 4-quarks has 0 hyper-charge,  4 Higgs insertions are needed to make it $U(1)_Y$ invariant, and there is a unique way to contract the $SU(2)_L$ indices to make it invariant: 
\bea
 {\rm LM24} &=&  i \sigma^{(2)}_{a b}  i \sigma^{(2)}_{de} H^*_a H^*_c  H^*_d H^*_f 
	(\overline{Q}_b \gamma^\mu Q _c ) 
        ( \overline{Q} _e \gamma^\mu Q_f) 
       (\overline{e}_R   e^C_R ) \\
       & \rightarrow & \frac{1}{4} v^4 (\overline{u}_L \gamma^\mu d_L)(\overline{u}_L \gamma_\mu d_L) (\overline{e}_R e^C)
\eea
This operator only contributes to SS dilepton. This is the operator whose 4-quark matrix element can be related using $SU(3)_L \times SU(3)_R$ flavor symmetry to the amplitude for $K \rightarrow \pi \pi$ \cite{Savage:1998yh}. 
In the effective Lagrangian after electroweak symmetry breaking the coefficient of this operator is suppressed by $v^4 /\Lambda^4$. 

The results of this Section are summarized in Table \ref{not-so-BigTable3}.

\begin{table*}
\centering
\resizebox{\textwidth}{!}{
\begin{tabular}{c|c|c|c|c|c|c|}
\hline \hline
\multirow{2}{*}{operator} & \multirow{2}{*}{content} & \multicolumn{3}{c|}{hadron collider signatures} & \multirow{2}{*}{Low Energy} & \multirow{2}{*}{$\chi$PT $(\pi \pi)$} \\ 
\cline{3-5}
& & \multicolumn{1}{p{2cm}|}{same-sign dilepton} &   \multicolumn{1}{c|}{$e$+MET} &  \multicolumn{1}{p{2.5cm}|}{dijet+ MET} &  & \\
\hline 
\multicolumn{7}{c}{dimension 11} \\ 
\hline
 & & & & & & \\       
LM20 & $  i\sigma^{(2)}_{ae}i \sigma^{(2)}_{cd} H^*_{e} H^*_{b}  
	(\overline{Q}_{a} \gamma^\mu Q_{b})
        (\overline{Q}_{c} d_R)
        (\overline{\ell}_{d} \gamma_\mu e^C_R) $ & $\surd$ &  $\surd$   &  $\ddot{\frown}$  & ${\cal O}^\mu_{LLLR}\otimes (LR) $  & - \\ 
        & & & &  & & \\
 LM21 & $  i\sigma^{(2)}_{ae}i \sigma^{(2)}_{cd} H^*_{e} H^*_{b}  
	(\overline{Q}_{a} \gamma^\mu \lambda^A Q_{b})
        (\overline{Q}_{c} \lambda^A d_R)
        (\overline{\ell}_{d} \gamma_\mu e^C_R) $ & $\surd$ &  $\surd$   &  $\ddot{\frown}$  & ${\cal O}^{\lambda \mu}_{LLLR}\otimes (LR) $  & - \\ 
        & & & &  & & \\ 
LM22a & $   i\sigma^{(2)}_{ab}  H^*_b H^*_{c}  
	(\overline{Q}_{a} \gamma^\mu Q_{c}) 
        ( \overline{u}_R Q_{d})
        (\overline{\ell}_{d}  \gamma_\mu  e^C_R )   $ & $\surd$ & $\surd$   &  $\ddot{\frown}$  & ${\cal O}^\mu_{LLRL}\otimes (LR) $  & - \\
        & & & & &  & \\
LM22b  & $   i \sigma^{(2)}_{ab} H^*_{b} H^*_{d}  
	( \overline{Q}_{a} \gamma^\mu  Q_{c})
        (\overline{u}_R Q_{d}) 
        ( \overline{\ell} _{c} \gamma_\mu e^C_R)  $ & $\surd$ & $\surd$   &  $\ddot{\frown}$  & ``\hbox{same as LM22a}"   & - \\ 
        & & & &  &  & \\
LM22c  & $  i \sigma^{(2)}_{ab} H^*_{c} H^*_{d}  
	( \overline{Q}_{a} \gamma^\mu  Q_{c})
        (\overline{u}_R Q_{d}) 
        ( \overline{\ell} _{b} \gamma_\mu  e^C_R)  $ & $\surd$ & $\surd$   &  $\ddot{\frown}$  & ``\hbox{same as LM22a}"  & -  \\ 
        & & & & & & \\
LM23a & $   i\sigma^{(2)}_{ab}  H^*_b H^*_{c}  
	(\overline{Q}_{a} \gamma^\mu \lambda^A Q_{c}) 
        ( \overline{u}_R \lambda^A Q_{d})
        (\overline{\ell}_{d}  \gamma_\mu  e^C_R )   $ & $\surd$ & $\surd$   &  $\ddot{\frown}$  & ${\cal O}^{\lambda \mu}_{LLRL}\otimes (LR) $  & - \\
        & & & & &  & \\
LM23b  & $   i \sigma^{(2)}_{ab} H^*_{b} H^*_{d}  
	( \overline{Q}_{a} \gamma^\mu \lambda^A Q_{c})
        (\overline{u}_R \lambda^A Q_{d}) 
        ( \overline{\ell} _{c} \gamma_\mu e^C_R)  $ & $\surd$ & $\surd$   &  $\ddot{\frown}$  & ``\hbox{same as LM23a}"   & - \\ 
        & & & &  &  & \\
LM23c  & $  i \sigma^{(2)}_{ab} H^*_{c} H^*_{d}  
	( \overline{Q}_{a} \gamma^\mu \lambda^A Q_{c})
        (\overline{u}_R \lambda^A Q_{d}) 
        ( \overline{\ell} _{b} \gamma_\mu  e^C_R)  $ & $\surd$ & $\surd$   &  $\ddot{\frown}$  & ``\hbox{same as LM23a}"  & -  \\ 
        & & & & & & \\ 
 \hline
 \hline 
 \end{tabular}}
\caption{Table of dimension-11 electroweak invariant operators contributing to $0 \nu \beta \beta$ decay and hadron collider processes. 
 After restricting the Higgs field to its vev, these operators do not reproduce any of the operators appearing in Tables \ref{BigTable1} or \ref{BigTable2}. 
Same notation as Table \ref{BigTable1}. All the quark operators appearing in this Table are vector. A `-' indicates the operator does not contribute to NNLO order.
\label{BigTable3}}
\end{table*}
 
\begin{table*}
\centering
\resizebox{\textwidth}{!}{
\begin{tabular}{c|c|c|c|c|c|c|}
\hline \hline
\multirow{2}{*}{operator} & \multirow{2}{*}{content} & \multicolumn{3}{c|}{hadron collider signatures} & 
 \multirow{2}{*}{Low Energy} & \multirow{2}{*}{$\chi$PT $(\pi \pi)$}  \\ 
\cline{3-5}
& & \multicolumn{1}{p{2cm}|}{same-sign dilepton} &   \multicolumn{1}{c|}{$e$+MET} &  \multicolumn{1}{p{2.5cm}|}{dijet+ MET} & & \\
\hline 
 \multicolumn{7}{c}{dimension 13} \\ 
\hline 
& & & & & & \\ 
LM24 & $  i \sigma^{(2)}_{a b} i  \sigma^{(2)}_{de} H^*_a H^*_c  H^*_d H^*_f 
	(\overline{Q}_b \gamma^\mu Q _c ) 
        ( \overline{Q} _e \gamma^\mu Q_f) 
       (\overline{e}_R   e^C_R ) $ & $\surd$ &   $\ddot{\frown}$ &  $\ddot{\frown}$  & ${\cal O}_{3L}\otimes (RR)$ & NNLO \\ 
       & & & & & & \\     
      \hline \hline
\end{tabular}}
\caption{Electroweak invariant dimension-13 operator contributing to $0 \nu \beta \beta$ decay and hadron collider processes. After restricting the Higgs field to its vev, this operator does not reproduce any of the operators appearing in Tables \ref{BigTable1}, 
\ref{BigTable2} or \ref{BigTable3}. Same notation as Table \ref{BigTable1}.
\label{not-so-BigTable3}}
\end{table*}

\section{Conclusions}
\label{conclusions}

In this paper we enumerate those short distance $\Delta L=2$ violating, baryon conserving, dimension-9 operators involving 4-quarks and two charged leptons that can contribute to a neutrinoless double $\beta$ decay signal. Compared to previous results \cite{Babu:2001ex,Pas:2000vn,Prezeau:2003xn}, here we impose electroweak invariance on the operators and 
determine their possible Lorentz and color structures. At the level of color and electromagnetic invariance only, here we find a minimal basis of 24 dimension-9 operators, which cannot be reduced any further through any combination of Lorentz or color Fierz transformations. The requirement of electroweak invariance is found to imply a set of 11 dimension-9 operators, a set much smaller than is allowed by electromagnetic invariance alone.  Those operators that do not occur at dimension-9 because they violate electroweak invariance are found to first appear at dimension-11 and, for one such operator, dimension-13.
Electroweak invariance implies additional collider signatures of such operators in final states involving neutrinos, which could in principle be detected, but whether that is possible in practice deserves further study. These results are summarized in Table \ref{BigTable1}.

We also set up a systematic mapping of the general set of 4-quark operators relevant for neutrinoless double beta decay onto chiral operators defined in chiral perturbation theory. Specifically, the chiral operators considered here involve pions coupled to 0, 2 or 4 nucleons. It has been known that of these chiral operators, those that couple the two leptons to two pions can lead to an enhanced decay rate compared to couplings between the leptons and four nucleons or with two nucleons and a pion, due to the long-range feature of the pion field. The reader is referred again to Fig. \ref{fig:chiPT}. Because of this possible enhancement, in this paper we determine the mapping of the 4-quark operators onto two pions at leading chiral order, confirming the leading order results found in Ref. \cite{Prezeau:2003xn}. For the phenomenology of the neutrinoless double beta decay rate, an important finding of Ref. \cite{Prezeau:2003xn} and confirmed here is that not all hadronic operators are found to have LO couplings to two-pions. 
These results are summarized in Tables \ref{table:scalar} and \ref{table:vector}, and the last column of Table \ref{BigTable1}.

It is hoped that the results presented here provide a systematic basis for future explorations of the effects of short distance $\Delta L=2$ processes on neutrinoless double beta decay. These directions include determining the complementarity between hadron collider and neutrinoless double beta decay bounds on such operators, as done in Ref. \cite{Peng:2015haa} for a specific model. Several physical effects must also be put together 
in order to perform accurate predictions of the neutrinoless double beta decay rate and to relate constraints from the LHC and neutrinoless double beta decay experiments. These inputs are: the QCD and electroweak renormalization effects which in general mix such operators; 
the lattice QCD matrix elements of such operators between pions and nucleons; and 
the mapping of the full set of neutrinoless double $\beta$ operators onto chiral operators. 

\acknowledgments
The author thanks Vincenzo Cirigliano, Claude Duhr, Chris Lee, Tao Peng, Michael Ramsey-Musolf, Martin Savage, and Peter Winslow for discussions. 
The author is grateful to V. Cirigliano,  W. Dekens, E. Mereghetti, and B. Tiburzi for discussions on the operator basis.
The author especially thanks Vincenzo Cirigliano for comments on an earlier draft of this manuscript. The author also thanks Bhupal Dev and Shao-Feng Ge for comments on an earlier version of this manuscript. 
This work has been supported by the U.S. Department of Energy at Los 
Alamos National Laboratory under Contract No. DE-AC52-06NA25396. The author thanks the LANL LDRD program office for support of this work. The preprint number for this manuscript is LA-UR-16-23550.

\section{Appendix} 

\subsection{Charge Conjugation Notation} 

Charge conjugation is given here by $\psi^c \equiv i \gamma_2 \psi^*$, which is the same as the {\tt FeynRules} definition 
CC$[\psi]=C (\overline{\psi})^T$ with $C=i \gamma_2 \gamma_0$. We also denote $e^C_R \equiv (e_R)^c$ and $\ell^C \equiv (\nu_L~ l_L)^c$. 

\subsection{Complete Basis of Dimension 9 operators after electroweak symmetry breaking} 
\label{Appendix:complete-set}

In this Appendix, I enumerate all possible dimension-9 operators contributing to $\Delta L=2$ that are only $SU(3)_c \times U(1)_{em}$ invariant. 
Such a basis is relevant for both matching electroweak invariant operators to operators defined below the scale of electroweak symmetry breaking, and for matching onto the chiral effective field theory defined below the GeV scale. We also refer the reader to the more concise Appendix of Ref. \cite{Prezeau:2003xn}, that arrives at some of the same conclusions as presented here. 

The operators of interest involve four quarks and two leptons and will be a product of three spinor bilinears that are one of the two following forms 
\beq
(\overline{q}_{1} \Gamma_1 q_{2}) (\overline{q}_3 \Gamma_2 q_4) (\overline{e} \Gamma_3 e^c),
\label{4q-2l}
\eeq
or 
\beq
(\overline{q}_{1} \Gamma_4 q_{2}) (\overline{q}_3 \Gamma_5 e^c) (\overline{e} \Gamma_6 q_4),
\label{2q-2l-q}
\eeq
for some gamma matrices $\Gamma_{1-6}$ which are linear combinations of the sixteen gamma matrices $\Gamma^A=\{\mathbb{1}, \gamma^\mu, \sigma^{\mu \nu}, \gamma^\mu \gamma^5, i\gamma^5\}$ that are a complete basis for $4\times4$ matrices, normalized to tr$[\Gamma^A \Gamma_B]=4\delta^A_B$. 

In Eqns. (\ref{4q-2l}) and (\ref{2q-2l-q}) the two leptons are either together, or one is each with one quark. The second class of operators is redundant, for using the following generalized Fierz transformation
(with no sum over spinor indices $i, j$, $\overline{i}$, $\overline{j}$), 
\beq 
(\Gamma^A)_{\overline{i} j} \otimes (\Gamma_B)_{\overline{j} i} = \frac{1}{16}\sum_{X,Y}  \hbox{tr}[\Gamma^A \Gamma^Y \Gamma_B \Gamma_X] (\Gamma^X)_{\overline{i} i} \otimes (\Gamma_Y)_{\overline{j} j},
\eeq
one can put operators of the second class into operators from the first class. Here 
the sums for $X$ and $Y$ are each over the complete set of gamma matrices.  As an aside, one can show using the properties $\{\Gamma^Z, \Gamma^{Z^\prime}\}_{\pm}=0$ that if $A=B$ (no sum) then only diagonal terms $X=Y$ contribute to the right-side of the above relation, and then the above formula reduces to the ``standard" Fierz table appearing in textbooks (see for e.g. \cite{Itzykson:1980rh}). 

The quarks are made $SU(3)_c$ invariant by contracting a quark with an anti-quark, giving two possibilities, or using the $SU(3)_c$ generators to form a singlet out of two color-octet operators. These three options are not independent, because of the $SU(N)$ Fierz identity for the fundamental representation 
\beq
\delta_{\alpha \sigma} \delta_{\rho \beta} = \frac{1}{N} \delta_{\alpha \beta} \delta_{\rho \sigma} +  \lambda^A_{\alpha \beta} \lambda^A _{\rho \sigma},
\label{color-Fierz}
\eeq
where we normalize the generators as Tr$[\lambda^A \lambda^B]=\delta^{AB}$, $\alpha, \beta, \rho, \sigma =1,..., N$, and we sum over 
$A=1,...,N^2-1$.

The next step is to enumerate possible $\Gamma$ structures for the lepton bilinear and the two quark bilinears. 
Since $\overline{\psi} \gamma^\mu \psi^c$=$\overline{\psi} \sigma^{\mu \nu} \psi^c$=$\overline{\psi} \sigma^{\mu \nu} \gamma_5 \psi^c$=0, the only same-flavor (SF) lepton bilinears are $\Gamma^C=\overline{\psi} \psi^c$, $\overline{\psi} \gamma_5 \psi^c$, and $\overline{\psi} \gamma^\mu \gamma_5 \psi^c$.  If we work with fields having definite chirality, then these three possibilities correspond to $\overline{e}_{L/R} e^c_{L/R}$ and $\overline{e}_{L} \gamma^\mu e^c_{R}-\overline{e}_{R} \gamma^\mu e^c_{L} $.

Next consider the two quark bilinears and work in the basis of quark fields with definite chirality. Allowing for all possible chiralities for the four quarks, the possible tensor products are either scalar or tensor: 
\begin{itemize} 
\item  $\mathbb{1} \otimes \mathbb{1}$, 
\item $\sigma^{\mu \nu} \otimes  \sigma^{\rho \sigma}$,
\item $\gamma^\mu \otimes \gamma^\nu$,
\end{itemize} 
or vector: 
\begin{itemize}  
\item $\mathbb{1} \otimes \gamma^\mu$, 
\item $\gamma^\mu \otimes \sigma^{\rho \sigma}$. 
\end{itemize} 
We consider these in turn.

\subsubsection{$\mathbb{1} \otimes \mathbb{1}$}

These operators are of the form 
\beq 
(\overline{q}_{L/R} q_{R/L}) (\overline{q}_{L/R} q_{R/L}), (\overline{q}_{L/R} q_{R/L}) (\overline{q}_{R/L} q_{L/R}) \nonumber
\eeq
where for each operator both types of color contractions must be considered. In the effective Lagrangian these operators are multiplied by $\overline{e}e^c$ or $\overline{e}\gamma_5 e^c$.

A Fierz transformation relates the last operator to other operators above, 
\bea
{\cal O}^\prime_{2LR}= (\overline{q}^\alpha_{L} q_{\alpha R}) (\overline{q}^\beta_{R} q_{\beta L}) &=& -\frac{1}{4} [ (\overline{q}^\alpha_L \gamma^\mu q_{\beta L})(\overline{q}^\beta_R \gamma^\mu q_{\alpha R}) + 
(\gamma^\mu \rightarrow \gamma^5 \gamma^\mu)] \nonumber \\
&=& -\frac{1}{2} [ (\overline{q}^\alpha_L \gamma^\mu q_{\beta L})(\overline{q}^\beta_R \gamma^\mu q_{\alpha R})]  \nonumber \\
&=&- \left[\frac{1}{6}  (\overline{q}^\alpha_L \gamma^\mu q_{\alpha L})(\overline{q}^\beta_R \gamma^\mu q_{\beta R}) \right. \nonumber \\ 
& & \left. +
\frac{1}{2}  (\overline{q}^\alpha_L \gamma^\mu \lambda^A q_{\alpha L})(\overline{q}^\beta_R \gamma^\mu \lambda^A q_{\beta R}) \right] \nonumber \\
&=& -\frac{1}{6} {\cal O}_{1LR} - \frac{1}{2} {\cal O}^\lambda_{1LR}~. \nonumber
\eea
Of the operators ${\cal O}_{1LR}$, ${\cal O}^{\lambda}_{1LR}$ and ${\cal O}^\prime_{2LR}$ only two are independent, out of which we choose the first two. 
Similarly, the other color singlet operator is not independent of ${\cal O} _{1LR}$:
\bea
{\cal O}^{\prime \prime}_{2LR}=(\overline{q}^\alpha_{L} q_{\beta R}) (\overline{q}^\beta_{R} q_{\alpha L}) &=& -\frac{1}{4} [ (\overline{q}^\alpha_L \gamma^\mu q_{\alpha L})(\overline{q}^\beta_R \gamma^\mu q_{\beta R}) + 
(\gamma^\mu \rightarrow \gamma^5 \gamma^\mu)] \nonumber \\
&=& -\frac{1}{2} [ (\overline{q}^\alpha_L \gamma^\mu q_{\alpha L})(\overline{q}^\beta_R \gamma^\mu q_{\beta R})]  \nonumber \\
&=& - \frac{1}{2} {\cal O} _{1LR}~. \nonumber
\eea
This leaves two four-quark operators of the form 
\bea 
(\overline{q}_{L/R} q_{R/L}) (\overline{q}_{L/R} q_{R/L}) \nonumber
\eea
and two more with the $SU(3)_c$ generators inserted
\bea 
(\overline{q}_{L/R} \lambda^A q_{R/L}) (\overline{q}_{L/R}  \lambda^A q_{R/L}) ~.\nonumber
\eea
The first two operators are just ${\cal O}_{2LR}$ and ${\cal O}_{2RL}$, and the last two are
just ${\cal O}^\lambda_{2RL}$ and ${\cal O}^\lambda_{2LR}$. 

\subsubsection{$\sigma^{\mu \nu} \otimes  \sigma^{\rho \sigma}$}

We cannot contract with a SF dilepton, because $\overline{e} \sigma^{\mu \nu} e^c =0$. And there aren't the right number of Lorentz indices to form a Lorentz scalar by contracting with $\overline{e} \gamma^\mu \gamma_5 e^c$. That leaves contracting $\sigma^{\mu \nu} \otimes \sigma_{\mu \nu}$ and multiplying by $\overline{e} e^c$ or $\overline{e} \gamma_5 e^c$. 

One can show using a Fierz transformation that 
\bea 
(\overline{q}_L \sigma^{\mu \nu} q_R) (\overline{q}_R \sigma_{\mu \nu} q_L) &=&0, \nonumber\\
 (\overline{q}_L \sigma^{\mu \nu} \lambda^A q_R) (\overline{q}_R \sigma_{\mu \nu} \lambda^A q_L) &=&0 ~.\nonumber
\eea
Next, a Fierz transformation on the remaining $\sigma^{\mu \nu} \otimes \sigma_{\mu \nu}$ operators shows they can be expressed in terms of previously defined operators. Namely,  
\bea 
{\cal O}_{4RL} &=& (\overline{q}^{\alpha}_{R} \sigma^{\mu \nu} q_{\alpha L}) (\overline{q}^\beta_R \sigma_{\mu \nu} q_{\beta L}) \nonumber \\
 &=& -8 (\overline{q}^\alpha_{R} q_{\beta L }) (\overline{q}^\beta_{R} q_{\alpha L}) + 4  (\overline{q}^\alpha_{R} q_{\alpha L}) (\overline{q}^\beta_{R} q_{\beta L}) \nonumber \\
{\cal O}_{4LR} &=& (\overline{q}^{\alpha}_{R} \sigma^{\mu \nu} q_{\beta L}) (\overline{q}^\beta_R \sigma_{\mu \nu} q_{\alpha L}) \nonumber \\
 &=& -8 (\overline{q}^\alpha_{R} q_{\alpha L}) (\overline{q}^\beta_{R} q_{\beta L}) + 4  (\overline{q}^\alpha_{R} q_{\beta L}) (\overline{q}^\beta_{R} q_{\alpha L}) \nonumber
 \eea
In short, in a chiral basis, all operators of the form $\sigma^{\mu \nu} \otimes  \sigma^{\rho \sigma}$ either vanish or can be expressed in terms of previously defined operators. 

\subsubsection{$\gamma^\mu \otimes \gamma^\nu$} 
Here we have to contract the two $\gamma^\mu$'s with each other, since the alternative is to contract them with the lepton bilinear, but for SF leptons $\overline{e} \sigma^{\mu \nu} e^c=0$. Operators in this category are therefore of the form 
\begin{subequations}
\bea 
{\cal O}_{3L/3R} & \equiv &(\overline{q}_{L/R} \gamma^\mu q_{L/R}) (\overline{q}_{L/R} \gamma_\mu q_{L/R}), \\ 
{\cal O}_{1LR}  & \equiv &(\overline{q}_{L/R} \gamma^\mu q_{L/R}) (\overline{q}_{R/L} \gamma_\mu q_{R/L}), \\
{\cal O}^\lambda_{1LR}  & \equiv &(\overline{q}_{L/R} \lambda^a \gamma^\mu q_{L/R}) (\overline{q}_{R/L} \lambda^a \gamma_\mu q_{R/L})
\eea
\end{subequations}
A standard Fierz transformation shows that 
\bea 
(\overline{u}^\alpha_L \gamma^\mu d_{\beta L})(\overline{u}^\beta_L \gamma^\mu d_{\alpha L})=
(\overline{u}^\alpha_L \gamma^\mu d_{\alpha L})(\overline{u}^\beta_L \gamma^\mu d_{\beta L}) 
\eea 
and similarly for $L \rightarrow R$, so that ${\cal O}_{3L/3R}$ and ${\cal O}^\lambda_{3L/3R}$ are not independent. We choose ${\cal O}_{3L, 3R}$, ${\cal O}_{1LR}$,  and ${\cal O}^\lambda_{1LR}$  to be part of the minimal basis. 

\subsubsection{$\mathbb{1} \otimes \gamma^\mu$}
Here there are eight operators,  
\begin{subequations}
\bea 
(\overline{q}^\alpha_{L/R} \gamma^\mu q_{\alpha L/R})(\overline{q}^\beta_L q_{\beta R}), & & (\overline{q}^\alpha_{L/R} \gamma^\mu q_{\alpha L/R}) (\overline{q}^\beta_R q_{\beta L}), \\
(\overline{q}^\alpha_{L/R} \gamma^\mu q_{\beta L/R})(\overline{q}^\beta_L q_{\alpha R}), & & (\overline{q}^\alpha_{L/R} \gamma^\mu q_{\beta L/R}) (\overline{q}^\beta_R q_{\alpha L})
\label{1-gammamu}
\eea 
\end{subequations}
In the effective Lagrangian these operators are multiplied by $\overline{e} \gamma^\mu \gamma_5 e^c$. 
These quark operators are just the operators previously defined in Eqns. (\ref{vectorOLLLR}--\ref{vectorOlRRRL}), after color-Fierzing the operators in the second line above. 

\subsubsection{$\gamma^\mu \otimes \sigma^{\rho \sigma}$} 
The only non-vanishing contraction with a lepton bilinear is  $(\gamma_\nu \otimes \sigma^{\nu \mu}) (\overline{e} \gamma_5 \gamma_\mu e^c$). A generalized Fierz transformation however relates all these operators to those appearing in Eqn. (\ref{1-gammamu}). To see that it is easier to work in two-component notation. In Appendix B of the review by Dreiner, Haber, and Martin \cite{Dreiner:2008tw}, one finds the 21 generalized Fierz identities expressed in two-component notation. Four of these relations are relevant for this class of operators, namely Eqns B.1.8-B.1.11 from that reference, 
\begin{subequations}
\bea 
\delta_\alpha^{~\beta} \sigma^{\mu}_{\gamma \dot{\alpha}} &=& \frac{1}{2} \sigma^\mu_{\alpha \dot{\alpha}} \delta^{~\beta}_\gamma- i \sigma_{\nu \alpha \dot{\alpha}} (\sigma^{\mu \nu})^{~\beta}_\gamma \\
\delta_\alpha^{~\beta} \overline{\sigma}^{\mu \dot{\beta} \gamma} &=& \frac{1}{2} \delta^{~\gamma}_\alpha \overline{\sigma}^{\mu \dot{\beta} \beta}+i (\sigma^{\mu \nu})^{~\gamma}_\alpha \overline{\sigma}^{\dot{\beta} \beta}_\nu \\
\delta^{\dot{\alpha}}_{~\dot{\beta}} \sigma^\mu _{\beta \dot{\gamma}} &=& \frac{1}{2} \delta^{\dot{\alpha}}_{~\dot{\beta}} \sigma^\mu _{\beta \dot{\beta}}+i (\overline{\sigma}^{\mu \nu})^{\dot{\alpha}}_{~\dot{\gamma}} \sigma_{\nu \beta \dot{\beta}}\\
\delta^{\dot{\alpha}}_{~\dot{\beta}} \overline{\sigma}^{\mu \dot{\gamma} \alpha} &=& \frac{1}{2} \overline{\sigma}^{\mu \dot{\alpha} \alpha} \delta^{\dot{\gamma}}_{~ \beta} 
-i \overline{\sigma}^{\dot{\alpha} \alpha}_{\nu} (\overline{\sigma}^{\mu \nu})^{\dot{\gamma}}_{~ \dot{\beta}}
\eea  
\end{subequations}
For a given color-ordering and helicity structure of an operator $\gamma_\nu \otimes \sigma^{\nu \mu}$, these Fierz identities relate that operator to the two possible color ordering of operators of the same helicity structure and type $\mathbb{1} \otimes \gamma^\mu$. As a result, in a chiral basis, all operators in this class can be eliminated.

\subsection{Relation of operator basis to prior literature} 
\label{app:relationtootherdefs}

In this work we organize the 4-quark operators by their transformation properties under chirality, which allows for an easier identification of their completion to $SU(2)_L \times U(1)_Y$ invariant operators and of their mapping onto operators in the chiral theory. Ref. 
\cite{Prezeau:2003xn} organizes the 4-quark operators by their parity transformation properties, and here we briefly make contact between the two notations. That reference defines nine 4-quark operators 
\begin{subequations}
\begin{eqnarray}
{\cal O}^{ab}_{1+} & = & (\overline{q}_L \tau^a \gamma^\mu q_L) 
 (\overline{q}_R \tau^b \gamma_\mu q_R) ,
  \label{O:1} \\ 
{\cal O}^{ab}_{2\pm} & = & (\overline{q}_R \tau^a  q_L) 
 (\overline{q}_R \tau^b  q_L) \pm (\overline{q}_L \tau^a  q_R) 
 (\overline{q}_L \tau^b  q_R)  ,
  \label{O:2}
  \\ 
{\cal O}^{ab}_{3\pm} & = & (\overline{q}_L \tau^a \gamma^\mu  q_L) 
 (\overline{q}_L \tau^b \gamma_\mu q_L) \pm 
(\overline{q}_R \tau^a \gamma^\mu  q_R) 
 (\overline{q}_R \tau^b \gamma_\mu q_R),
  \label{O:3}
  \\ 
{\cal O}^{ab,\mu}_{4\pm} & = & (\overline{q}_L \tau^a \gamma^\mu  q_L \mp
 \overline{q}_R \tau^a \gamma_\mu q_R)  
(\overline{q}_L \tau^b q_R - 
 \overline{q}_R \tau^b q_L),
  \label{O:4}
 \\ 
{\cal O}^{ab,\mu}_{5\pm} & = & (\overline{q}_L \tau^a \gamma^\mu  q_L \pm
 \overline{q}_R \tau^a \gamma_\mu q_R)  
(\overline{q}_L \tau^b q_R + 
 \overline{q}_R \tau^b q_L)~.
 \label{O:5}
\end{eqnarray}
\end{subequations}

I find that for scalar operators:
\begin{subequations}
\bea
{\cal O}_{1LR} &=& (\overline{q}_L \gamma^\mu \tau^+ q_L) (\overline{q}_R \gamma_\mu \tau^+ q_R), \\
	&=& {\cal O}^{++}_{1+} ,\\ 
{\cal O}_{2RL} &=& (\overline{q}_R  \tau^+ q_L) (\overline{q}_R  \tau^+ q_L), \\
 &=& \frac{1}{2}({\cal O}^{++}_{2+}+{\cal O}^{++}_{2-} ) ,\\
{\cal O}_{2LR} & = &(\overline{q}_L  \tau^+ q_R) (\overline{q}_L  \tau^+ q_R), \\
&=& \frac{1}{2}({\cal O}^{++}_{2+}-{\cal O}^{++}_{2-} ) ~,\\
{\cal O}_{3L} &=& (\overline{q}_L \gamma^\mu \tau^+ q_L) (\overline{q}_L\gamma_\mu \tau^+ q_L) , \\
&=& \frac{1}{2}({\cal O}^{++}_{3+}+{\cal O}^{++}_{3-} ) ,\\ 
{\cal O}_{3R} &= &(\overline{q}_R \gamma^\mu \tau^+ q_R) (\overline{q}_R\gamma_\mu \tau^+ q_R),\\
&=&\frac{1}{2}({\cal O}^{++}_{3+}-{\cal O}^{++}_{3-} ) , 
\eea
\end{subequations}
and for vector operators: 
\begin{subequations}
\bea
{\cal O}^\mu_{LLLR} &=& (\overline{q}_L \gamma^\mu \tau^+ q_L)(\overline{q}_L \tau^+ q_R), \\
&=& \frac{1}{4}({\cal O}^\mu_{4+} + {\cal O}^\mu_{4-} +{\cal O}^\mu_{5+} +{\cal O}^\mu_{5-}), \\
{\cal O}^\mu_{RRLR} &=& (\overline{q}_R \gamma^\mu \tau^+ q_R)(\overline{q}_L \tau^+ q_R), \\
&=& \frac{1}{4}(-{\cal O}^\mu_{4+} - {\cal O}^\mu_{4-} +{\cal O}^\mu_{5+} +{\cal O}^\mu_{5-}), \\
{\cal O}^\mu_{LLRL} &=& (\overline{q}_L \gamma^\mu \tau^+ q_L)(\overline{q}_R \tau^+ q_L), \\
&=& \frac{1}{4}(-{\cal O}^\mu_{4+} + {\cal O}^\mu_{4-} +{\cal O}^\mu_{5+} -{\cal O}^\mu_{5-}), \\
{\cal O}^\mu_{RRRL} &=& (\overline{q}_R \gamma^\mu \tau^+ q_R)(\overline{q}_R \tau^+ q_L), \\ 
&=& \frac{1}{4}({\cal O}^\mu_{4+} -{\cal O}^\mu_{4-} +{\cal O}^\mu_{5+} -{\cal O}^\mu_{5-})~. 
\eea
\end{subequations}

In addition to these operators, I also find additional operators - all involving color-octets - to be part of the minimal basis. These are the following 3 scalar operators - ${\cal O}^{\lambda}_{1LR}$, ${\cal O}^\lambda_{2LR}$, and ${\cal O}^\lambda_{2RL}$ - and the following 4 vector operators ${\cal O}^{\lambda \mu}_* $, where $*=LLLR,RRLR,LLRL$ or $RRRL$.

\bibliographystyle{JHEP}
\bibliography{0nubbbib}

 \end{document}